\documentclass{elsart}
\usepackage{amsmath,amssymb}
\usepackage[dvips]{graphicx}

\newcommand{\rmd}{\mathrm{d}}
\newcommand{\rme}{\mathrm{e}}
\newcommand{\rmi}{\mathrm{i}}
\newcommand{\sgn}{\mathrm{sgn}}
\newcommand{\tr}{\mathrm{tr}}
\newcommand{\Dphi}{\Delta\phi}
\newcommand{\bJ}{\boldsymbol{J}}
\newcommand{\bA}{\boldsymbol{A}}
\newcommand{\bB}{\boldsymbol{B}}
\newcommand{\bE}{\boldsymbol{E}}

\newcommand{\bp}{\boldsymbol{p}}
\newcommand{\bq}{\boldsymbol{q}}
\newcommand{\llangle}{\langle\!\langle}
\newcommand{\rrangle}{\rangle\!\rangle}
\renewcommand{\P}{\mathcal{P}}
\newcommand{\CP}{\mathcal{CP}}
\newcommand{\W}{\mathcal{W}}
\newcommand{\MeV}{\;\text{MeV}}
\newcommand{\GeV}{\;\text{GeV}}
\newcommand{\fm}{\;\text{fm}}
\newcommand{\feyn}[1]{
  \setbox0=\hbox{\ensuremath{#1}}
  \hbox to\wd0{\hbox to0pt{\hbox to\wd0{\hss/\hss}\hss}\box0}}

\begin{document}

\begin{flushright}
\vspace*{-1cm}
{\small\sf BNL-90863-2009-JA\\YITP-09-60}
\vspace{1cm}
\end{flushright}

\begin{frontmatter}

\title{Electric-current Susceptibility\\
       and the Chiral Magnetic Effect}

\author[Yukawa]{Kenji Fukushima,}
\author[BNL]{Dmitri E.\ Kharzeev} and
\author[Frankfurt]{Harmen J.\ Warringa}
\address[Yukawa]{Yukawa Institute for Theoretical Physics,
Kyoto University, Kyoto, Japan}
\address[BNL]{Department of Physics,
Brookhaven National Laboratory, Upton NY 11973, USA}
\address[Frankfurt]{Institut f\"ur Theoretische Physik,
Goethe-Universit\"at, Max-von-Laue-Stra\ss e 1, D-60438
Frankfurt am Main, Germany}

\begin{abstract}
 We compute the electric-current susceptibility $\chi$ of hot
 quark-gluon matter in an external magnetic field $B$.  The difference
 between the susceptibilities measured in the directions parallel and
 perpendicular to the magnetic field is ultraviolet finite and given
 by $\chi^\parallel-\chi^\perp = VTN_c\sum_f q_f^2 |q_f B|/(2\pi^2)$,
 where $V$ denotes the volume, $T$ the temperature, $N_c$ the number
 of colors, and $q_f$ the charge of a quark of flavor $f$.  This
 non-zero susceptibility difference acts as a background to the Chiral
 Magnetic Effect, i.e.\ the generation of electric current along the
 direction of magnetic field in the presence of topological charge.
 We propose a description of the Chiral Magnetic Effect that takes
 into account the fluctuations of electric current quantified by the
 susceptibility.  We find that our results are in agreement with
 recent lattice QCD calculations.  Our approach can be used to model
 the azimuthal dependence of charge correlations observed in heavy ion
 collisions.
\end{abstract}

\end{frontmatter}


\section{Introduction}
The strong $\CP$ problem -- the absence of $\P$ and $\CP$ violation in
strong interactions -- still stands as one of the fundamental puzzles
of contemporary physics.  The puzzle stems from the existence of
vacuum topological solutions in Quantum Chromo-Dynamics
(QCD)~\cite{Belavin:1975fg}.  Due to the axial anomaly~\cite{anomaly}
these topological solutions induce the non-conservation of
flavor-singlet axial current.  The resulting time-dependence of the
axial charge leads to the picture of the physical QCD vacuum
representing a Bloch-type superposition of an infinite number of
topologically distinct but degenerate in energy sectors connected by
the tunneling transitions -- so called
``$\theta$-vacuum''~\cite{Callan:1976je}.  Instead of keeping track of
this vacuum structure explicitly one can instead equivalently
reproduce its effect by adding to the QCD Lagrangian a new
``$\theta$-term'', $\theta \cdot Q$, where
$Q \equiv (g^2/16\pi^2)\,\tr F_{\mu\nu}\widetilde{F}^{\mu\nu}$ is the
density of topological charge.  Unless $\theta = 0$, this term
explicitly breaks $\P$ and $\CP$ symmetries of QCD.

Among the effects induced by the $\theta$-term is a generation of the
($\P$- and $\CP$-odd) electric dipole moments (e.d.m.'s) of hadrons.
The current experimental upper bound on the neutron's e.d.m. is
$|d_n|<2.9\times 10^{-26}\;e\cdot\text{cm}%
$~\cite{Baker:2006ts}~\footnote{The weak interactions induce $\CP$
  violation through the phases in the CKM matrix, but the resulting
  neutron's electric dipole moment is very small,
  $|d_n|\sim 10^{-32}\;e\cdot\text{cm}$.}.  Since inducing a non-zero
electric dipole moment requires flipping the chirality of the quark
that is achieved by the quark mass $m_q$ insertion, on dimensional
grounds one expects
$|d_n|\sim (e\,m_q/m_N^2)\,\theta \approx 10^{-16}\,%
\theta\;e\cdot\text{cm}$ where $m_N$ is a typical hadronic scale that
we have chosen as a nucleon mass.  This rough estimate combined with
the experimental measurement of neutron's e.d.m.\ leads to the bound
$|\theta| \leq 10^{-10}$.  (A more careful analysis yields a somewhat
tighter bound, $|\theta|<0.7\times 10^{-11}$~\cite{Kim:2008hd}.)  The
mechanism responsible for the unnatural smallness of the parameter
$\theta$ has not been established yet -- this represents the strong
$\CP$ problem.  One appealing explanation (that however has not been
confirmed yet) promotes $\theta$ into a dynamical axion
field~\cite{Wilczek:1977pj,Weinberg:1977ma} emerging as a
Nambu-Goldstone boson of an additional chiral
symmetry~\cite{Peccei:1977hh}; for a review, see
Ref.~\cite{Peccei:2006as}.

The strength of topological charge fluctuations in QCD vacuum is
quantified by topological susceptibility that is defined as a second
derivative of the QCD partition function with respect to $\theta$.  On
the other hand, the dependence of the partition function on $\theta$
at low energies is governed by chiral symmetry; as a result, the
topological susceptibility can be expressed in terms of $f_\pi$ and
$\langle\bar{q}q\rangle$~\cite{Leutwyler:1992yt,Vicari:2008jw}.  The
phase structure associated with finite $\theta$ (especially at
$\theta\approx\pi$)~\cite{Dashen:1970et} can also be investigated in
the framework of chiral effective models~\cite{Boer:2008ct}.  The
fluctuations of topological charge affect the mass spectrum and other
properties of hadrons.

At high temperature and at weak coupling, topological fluctuations in
QCD matter are enhanced due to the real-time ``sphalerons''  (akin to
the thermal activation
processes)~\cite{Klinkhamer:1984di,McLerran:1990de,Arnold:1996dy,%
Huet:1996sh,Bodeker:1998hm}.  At temperatures not far from the
deconfinement transition temperature $T_c$ the physics is essentially
non-perturbative, and the description of real-time dynamics has to
rely on numerical simulations or models.  It was envisioned early on
that $\P$- and $\CP$-odd condensates~\cite{Lee:1973iz} may develop
locally in dense hadronic matter~\cite{Lee:1974ma}.  It is thus
conceivable that $\P$ and $\CP$ symmetry may be violated
locally~\cite{Morley:1983wr}.

An explicit metastable solution describing a $\P$- and $\CP$-odd
``bubble'' in hot QCD matter at $T \sim T_c$ has been found by using a
chiral lagrangian description~\cite{Kharzeev:1998kz}.  This metastable
solution describes a local domain filled with the $\eta^\prime$
condensate (or, equivalently, characterized by a locally non-vanishing
$\theta$); it is somewhat analogous to the disoriented chiral
condensate~\cite{Anselm:1989pk} that is however unstable even
classically.  It has been proposed~\cite{Kharzeev:1999cz} that in
heavy ion collisions $\P$-odd bubbles would induce certain $\P$-odd
correlations in pion momenta, but the experimental study of these
correlations appeared challenging
experimentally~\cite{Voloshin:2000xf,Finch:2001hs}.

However some time ago it was proposed that the presence of magnetic
field and/or angular momentum in heavy ion collisions opens new
possibilities for the observation of $\P$- and $\CP$-odd effects.
Specifically, it was found that in the presence of magnetic field
and/or angular momentum the fluctuations of topological charge can be
observed directly since they lead to the separation of electric charge
along the axis of magnetic field due to the spatial variation of the
topological charge
distribution~\cite{Kharzeev:2004ey,Kharzeev:2007tn,Kharzeev:2009fn}
and to the generation of electric current due to the time dependence
of the topological charge
density~\cite{Kharzeev:2009fn,Kharzeev:2007jp,Fukushima:2008xe},
i.e.\ the ``Chiral Magnetic Effect'' (CME).

The experimental observable that is sensitive to this locally $\P$-
and $\CP$-odd charge separation in heavy ion collisions has been
proposed in Ref.~\cite{Voloshin:2004vk}.  The preliminary data from
STAR Collaboration at Relativistic Heavy Ion Collider at
Brookhaven~\cite{Selyuzhenkov:2005xa,Voloshin:2008jx} indicated the
presence of the charge-dependent azimuthal correlations, with the
magnitude consistent with the early rough
estimate~\cite{Kharzeev:2004ey}.  Recently, STAR Collaboration
presented the conclusive observation of charge-dependent azimuthal
correlations~\cite{:2009uh} possibly resulting from the ($\P$- and
$\CP$-odd) charge separation with respect to the reaction plane.
Because the sign of charge asymmetry is expected to fluctuate from
event to event, these measurements are performed on the event-by-event
basis.  The experimental observable~\cite{Voloshin:2004vk} measures
the strength of charge asymmetry fluctuations; even though it is
sensitive to parity-violating effects, it is $\P$- and $\CP$-even and
so ``conventional'' backgrounds have to be carefully studied.  None of
the existing event generators (such as MEVSIM, UrQMD, and HIJING) can
reproduce the observed effect~\cite{:2009uh} although the search for
other possible explanations of course has to continue; see
e.g.~\cite{Wang:2009kd}.  At the same time, it is of paramount
importance to establish a firm theoretical framework allowing for a
quantitative study of $\P$- and $\CP$-odd effects in heavy ion
collisions.

The STAR result has already excited significant interest, and a number
of recent theoretical studies address in detail the physics of the
Chiral Magnetic Effect.  The first lattice study of CME has been
performed by the ITEP lattice gauge group~\cite{Buividovich:2009wi} in
quenched QCD.\ \ Recently, the Connecticut
group~\cite{Abramczyk:2009gb} performed the first study of CME in full
QCD with $(2+1)$ light-quark flavors in the domain-wall formulation.
These lattice studies provide an important confirmation of the
existence of CME, but also highlight a need for a quantitative
theoretical understanding of the involved non-perturbative phenomena.
In this paper we will attempt to reproduce some of the lattice results
in an analytical approach.

The behavior of CME at strong coupling in the Sakai-Sugimoto model and
related theories has been explored through the AdS/CFT correspondence in
Refs.~\cite{Lifschytz:2009sz,Yee:2009vw,Rebhan:2009vc,Sahoo:2009yq,%
D'Hoker:2009bc}.  Some of these results at present are under
discussion -- for example, while Yee finds in Ref.~\cite{Yee:2009vw}
that the magnitude of CME at strong coupling is not modified relative
to the weak coupling case, the authors of Ref.~\cite{Rebhan:2009vc}
argue that the effect disappears in the strong coupling limit.  In
Ref.~\cite{Nam:2009jb} the CME at low temperatures has been studied
using the instanton vacuum model.  The electric dipole moment of QCD
vacuum in the presence of external magnetic field has been evaluated
in Ref.~\cite{Millo:2007im} using the chiral perturbation theory.
Extensive studies of the local $\P$- and $\CP$ violation in hot
hadronic matter and of the influence of magnetic field on the phase
transitions have been performed in Ref.~\cite{Fraga:2008qn}.  The
properties of hadronic matter in external magnetic fields have
attracted significant interest
recently~\cite{Cohen:2007bt,Menezes:2008qt}.  The
analytical~\cite{Kharzeev:2007jp} and
numerical~\cite{Skokov:2009qp,Okorokov:2009bf} evaluations of the
strength of magnetic field produced in heavy ion collisions yield the
values of the order of $|eB|\sim m_\pi^2$ for RHIC energies, making it
possible to study the interplay of QCD and QED phenomena.  The
violation of parity in cold and dense baryonic matter has been
investigated in Refs.~\cite{Andrianov:2007kz}.  A closely related
problem is the $\theta$-dependence in hadronic matter at finite baryon
and isospin densities~\cite{Metlitski:2005db} as well as at high
temperatures \cite{Parnachev:2008fy}.

Let us now discuss the physics of CME in more detail.  First, note
that a constant (homogeneous in space and time) $\theta\neq0$ cannot
induce the CME.\ \  This is because the $\theta$-term in the QCD
lagrangian represents a full divergence and thus cannot affect the
equations of motion.  The CME electric current can be induced only
when $\theta$ changes in time from a finite value in the metastable
state toward zero in the ground state;  a spatially inhomogeneous
$\theta$ distribution induces the electric dipole
moment~\cite{Kharzeev:2007tn,Kharzeev:2009fn}.  To give a quantitative
description of the CME, in Ref.~\cite{Fukushima:2008xe} we used the
chiral chemical potential $\mu_5$ that is proportional to the time
derivative of $\theta$: $\mu_5=\partial_0\theta/(2N_f)$.  The anomaly
relation allows us to find an exact expression for the induced current
which is proportional to $B$ and $\mu_5$.  In analogy to the ordinary
relation between the current and the electric field through the
electric conductivity, the coefficient of the current in response to
$B$ can be called the ``chiral magnetic conductivity.''  In
Ref.~\cite{Kharzeev:2009pj} the chiral magnetic conductivity was
computed as a function of the energy and the momentum at weak
coupling.  The chiral magnetic conductivity  was then evaluated by
holographic methods also at strong coupling~\cite{Yee:2009vw}.

The experimental observable studied by STAR
\cite{:2009uh,Voloshin:2009hr} measures the strength of event-by-event
fluctuations of charge asymmetry relative to the reaction plane,
i.e.\ along the direction of magnetic field (and of orbital momentum).
Therefore the quantity that needs to be evaluated theoretically is the
correlation function of electric charge asymmetry; under some
reasonable assumptions it can be related to the susceptibility of the
CME electric current.  This quantity has also been computed recently
in the lattice QCD
simulation~\cite{Buividovich:2009wi,Abramczyk:2009gb}.

In this paper we will compute the electric-current susceptibility in
an analytical approach.  We hope that this calculation will serve as a
step in a quantitative theoretical understanding of both experimental
and lattice results.  In our study we consider explicitly only the
quark sector; we justify this by the absence of perturbative
corrections to the axial anomaly that is driving the CME.\ \ This is
justified at weak coupling -- in other words, the CME current is not
perturbatively renormalized.  Whether or not the CME current undergoes
a non-perturbative renormalization, or whether it survives in the
strong coupling limit, is still an open
question~\cite{Yee:2009vw,Rebhan:2009vc}.  We will characterize the
real-time dynamics of topological charge in the system by a certain
distribution in the chiral chemical potential $\mu_5$.

This paper is organized as follows.  In Sec.~\ref{sec:observable} we
briefly review the experimental observable accessible at present in
heavy ion collisions and motivate the connection between this 
observable and the electric-current correlation function and
susceptibility.  The following Sec.~\ref{sec:susceptibility} is
divided into four subsections:  first, we re-derive our previous
result for the CME current~\cite{Fukushima:2008xe} and compute the
longitudinal susceptibility using the thermodynamic potential.
Calculating the transverse susceptibility is not so straightforward as
the longitudinal one, and we explain our method of computation in the
three following subsections.  We then proceed to Sec.~\ref{sec:data}
where we make a comparison of our results to the lattice QCD data and
provide the formulae that can be used in the description of
experimental data.  Finally, we summarize in Sec.~\ref{sec:summary}.


\section{Experimental Observables and Correlation Functions}
\label{sec:observable}
\begin{figure}[ht]
 \begin{center}
 \includegraphics[width=0.8\textwidth]{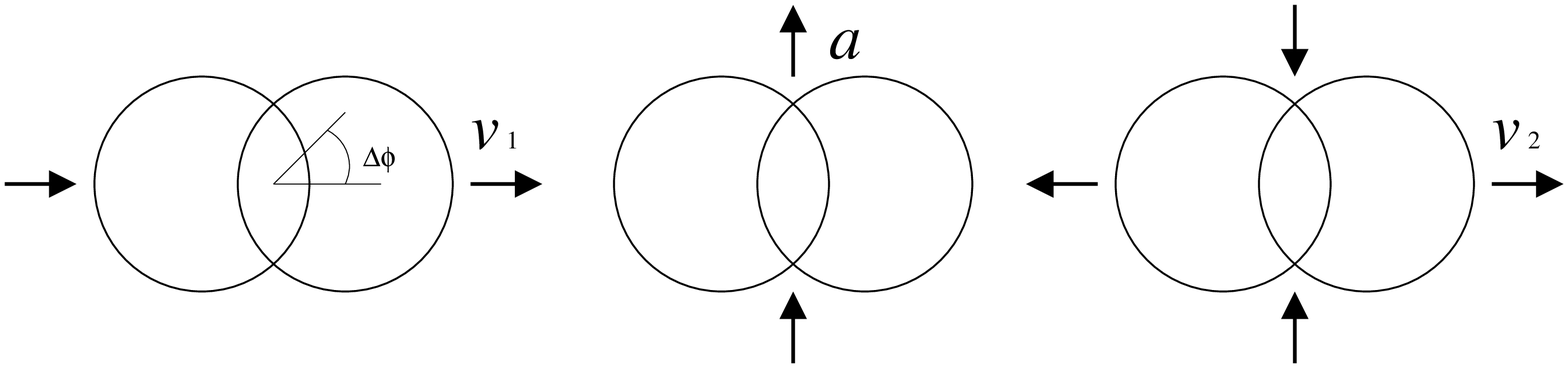}
 \end{center}
 \caption{Collision geometry and collective flows decomposed in
   Eq.~(\ref{eq:dNdphi}), of which $a_\pm$ is sensitive to $\P$-
   and $\CP$-odd effects.}
 \label{fig:flow}
\end{figure}

In heavy ion collisions at any finite impact parameter the
distribution of measured (charged) particles depends on the azimuthal
angle $\phi$.  The hadron multiplicity as a function of $\phi$ can be
decomposed into Fourier harmonics:
\begin{equation}
 \frac{\rmd N_\pm}{\rmd\phi} \propto 1+2v_{1\pm} \cos(\Dphi)
  +2a_\pm \sin(\Dphi) +2v_{2\pm} \cos(2\Dphi) + \cdots \,,
\label{eq:dNdphi}
\end{equation}
where $N_+$ and $N_-$ are the number of positively and negatively
charged particles respectively, and $\Dphi\equiv\phi-\Psi_{\text{RP}}$
is the angle relative to the reaction plane.  The coefficients $v_1$
and $v_2$ quantify the strength of ``directed'' and ``elliptic'' flows
respectively; they are not expected to depend on whether measured
particles are positively or negatively charged.  On the other hand,
the term proportional to $a_\pm$ is $\P$- and $\CP$- odd and describes
the charge separation relative to the reaction plane -- in other
words, as sketched in Fig.~\ref{fig:flow}, $a_\pm$ quantifies the
strength of the charge flow directed perpendicular to the reaction
plane.  Unlike $v_1$ and $v_2$, as we will see later, $a_+$ and $a_-$
depend on the charge carried by measured particles.  If the CME
electric current is directed upward (downward), we expect $a_+>0$
($a_-<0$); non-zero values of $a_+$ and $a_-$ indicate the presence of
$\P$- and $\CP$-odd effects.


We can carry out the decomposition~(\ref{eq:dNdphi}) for each event
and take the ensemble average over all the events.  We shall denote
this averaging procedure by $\llangle\cdots\rrangle$ throughout this
paper.  Because the topological excitations fluctuate not only locally
(point-by-point in space) but also globally (event-by-event),
$\llangle a_\pm\rrangle $ becomes zero and the symmetry is restored in
a sense of average.  It is necessary, therefore, to measure the
correlation functions
$\llangle a_+ a_+\rrangle$, $\llangle a_+ a_-\rrangle$, and
$\llangle a_- a_-\rrangle$, which are invariant under $\P$ and $\CP$
transformations and thus their ensemble average is non-vanishing.

The experimental observable sensitive to $\llangle a_\pm
a_\pm\rrangle$ was proposed by Voloshin~\cite{Voloshin:2004vk}:
\begin{equation}
 \llangle \cos(\Dphi_\alpha+\Dphi_\beta) \rrangle
  \equiv \biggl\langle\!\!\!\biggl\langle \frac{1}{N_\alpha N_\beta}
  \sum_{i=1}^{N_\alpha} \sum_{j=1}^{N_\beta}
  \cos(\Dphi_{\alpha,i}+\Dphi_{\beta,j}) \biggr\rangle\!\!\!
  \biggr\rangle \,.
\end{equation}
Here $\alpha$ and $\beta$ indicate either $+$ or $-$ charge of
measured particles, and the sum goes over all charged hadrons in a
given event.  This observable has an important property that becomes
clear when one explicitly isolates the terms $B_{\alpha\beta}$ driven
by fluctuating backgrounds~\cite{Voloshin:2004vk}:
\begin{align}
 \llangle \cos(\Dphi_\alpha+\Dphi_\beta) \rrangle
 &= \llangle \cos\Dphi_\alpha \cos\Dphi_\beta \rrangle
  -\llangle \sin\Dphi_\alpha \sin\Dphi_\beta \rrangle \notag\\
 &= \bigl( \llangle v_{1,\alpha} v_{1,\beta} \rrangle
  + B^{\text{in}}_{\alpha\beta} \bigr)  -  \bigl(
  \llangle a_\alpha a_\beta \rrangle + B^{\text{out}}_{\alpha\beta}
  \bigr) \,.
\label{eq:voloshin}
\end{align}
If the in-plane $B^{\text{in}}_{\alpha\beta}$ and out-of-plane
$B^{\text{out}}_{\alpha\beta}$ backgrounds are the same, they cancel
out and since for a symmetric heavy ion collision in a symmetric
rapidity cut $\llangle v_{1,\alpha}v_{1,\beta}\rrangle\approx 0$, one
can identify the above (\ref{eq:voloshin}) with
$-\llangle a_\alpha a_\beta\rrangle$ that serves as an order parameter
for $\P$- and $\CP$-odd effects.

Let us now take a further step to establish a link between the above
experimental observable and the theoretical computation.  In theory a
clear manifestation of the CME is spontaneous generation of an
electric current, $\bJ$, along the direction parallel to the applied
magnetic field, $\bB$~\cite{Kharzeev:2007jp,Fukushima:2008xe}.  We can
anticipate that this electric current of quarks, if $J_z>0$, would
induce $a_+>0$ and $a_-<0$ in Eq.~(\ref{eq:dNdphi}) for the observed
hadron distributions.  This is due to i) the quark-hadron duality and
ii) the rapid transverse expansion of the system that prevents the
charge asymmetry from being smeared out by thermal diffusion.  It
seems natural that a theoretical quantity relevant to
$\llangle a_\pm a_\pm\rrangle$ should be a correlation function or
fluctuation with respect to the electric current.  For one event it
would be legitimate to accept the following relations:
\begin{align}
 & \sum_{i=1}^{N_+} \cos\Dphi_{+,i} \;\approx\;
  -\sum_{i=1}^{N_-} \cos\Dphi_{-,i} \;\propto\; J_x \,, \\
 & \sum_{j=1}^{N_+} \sin\Dphi_{+,j} \;\approx\;
  -\sum_{i=1}^{N_-} \sin\Dphi_{-,i} \;\propto\; J_z \,.
\end{align}
Indeed, $\sum\cos\Dphi_{+,i}$, for example, counts an excess of
positively charged particles in the region near $\Dphi\approx 0$
(i.e.\ moving in the $x>0$ direction) as compared to that near
$\Dphi\approx \pi$ (i.e.\ moving in the $x<0$ direction).  Here we
defined $x$ and $z$ as the transverse and longitudinal coordinates,
respectively.  From now on, to make our notation more general, we will
use $J_\perp$ and $J_\parallel$ instead of specific coordinate
indices; namely $J_\perp$ is a current component perpendicular to
$\bB$ and $J_\parallel$ parallel to $\bB$.  We can then identify
\begin{equation}
 \cos(\Dphi_\alpha + \Dphi_\beta) \propto 
  \frac{\alpha\beta}{N_\alpha N_\beta} \bigl(
  J_\perp^2 - J_\parallel^2 \bigr) \,.
\label{eq:cosJ2}
\end{equation}
Note that we have neglected the screening (absorption) effect by the
medium for the moment; we will discuss the medium screening in more
detail in Sec.~\ref{sec:data}.

To connect these expressions to the experimental observables we should
further take the average $\llangle\cdots\rrangle$.  Computing this
event average would however require a complete real-time simulation
describing the dynamics of topological transitions in the produced
matter.  Since such a simulation goes beyond the scope of the present
work, here we will adopt the following \textit{working definition} for
the correlation function that is related to the experimental data.

\begin{enumerate}

\item  First, we identify the fluctuation measured on the
  event-by-event basis with the spatial correlation function obtained
  in the field theory calculation.  This is a common assumption used
  for example also in discussions on the QCD critical point
  search~\cite{Stephanov:1998dy}.  If the system's volume is
  sufficiently larger than the correlation length, this assumption
  should make sense.\\

\item Second, we compute the correlation function for a fixed value of
  the chiral chemical potential $\mu_5$ that induces a difference in
  number between right-handed and left-handed particles;
  $N_5=N_R-N_L$.\\

\item Third, we make a convolution of the results at fixed $\mu_5$
  with a weight function of the $\mu_5$ distribution.  We will adopt a
  Gaussian ansatz for the weight function; the dispersion is
  characterized by the rate of topological transitions depending on
  the system's temperature.

\end{enumerate}

One may wonder whether the above-mentioned procedure is a correct one,
or merely a convenient assumption.  Suppose that
$\langle J_\perp^2\rangle$ and $\langle J_\parallel^2\rangle$ were
computed from the first principles of quantum field theory avoiding
the steps (2) and (3) above, as is the case for the lattice QCD
simulations.  Then one might think that the results are to be
interpreted as quantities directly relevant to the experimental data.
We think however that the situation is not so simple.  The difficulty
lies in the different dynamics of topological excitations in Euclidean
and Minkowski space-times.  In the case of Euclidean space-time the
instantons are exponentially suppressed by the Boltzmann factor, while
in Minkowski space-time the sphalerons are not~\cite{Arnold:1987zg}.
Therefore, it is essential to take account of sphaleron-like effects
in real-time dynamics.  If we make use of the imaginary-time formalism
of the finite-temperature field theory, it would describe properly the
instanton transitions but would miss the unsuppressed sphaleron
contributions unless the analytical continuation is correctly
formulated.  Usually the analytical continuation from the imaginary-
to the real-time dynamics is not obvious though.  It would thus be
appropriate to solve the real-time dynamics e.g.\ by using the
spatially discretized
Hamiltonian~\cite{Grigoriev:1989ub,Ambjorn:1990wn,Moore:1997sn,%
Meggiolaro:1998bh,Kharzeev:2001ev} in the presence of external
magnetic field.  This is however quite difficult technically and
hence, in this work, we have employed a pragmatic approach assuming
that the computation of the spatial correlation function at finite
$\mu_5$ and the computation of the average over the distribution in
$\mu_5$ (that is governed by the real-time dynamics) are separable.
Such a separability would hold for example if the CME response of the
system to a finite $\mu_5$ were much faster than the dynamics
responsible for the generation of the distribution in $\mu_5$, and the
back-reaction of the CME current on topological transitions could be
neglected.  We think that this is a plausible assumption since the
generation of the CME current in response to a finite $\mu_5$ is not
damped by interactions with gluons, and at least at large $N_c$ the
back-reaction of the CME current on the topological dynamics of gluon
fields can be neglected.  In this approach, the properties of the
topological transitions in hot QCD medium such as the sphaleron rate
are incorporated through the weight function for the
$\mu_5$-distribution.

The correlation function, in general, has two contributions: one is
from a disconnected (tadpole-tadpole type) diagram and the other from
a connected diagram.  The former is related to the expectation value
of the induced current, and the latter is given by the corresponding
susceptibility.  Here we introduce the following notations:
\begin{equation}
 \langle J_\parallel^2 \rangle_{\mu_5}
  = \langle J_\parallel \rangle_{\mu_5}^2
  + \chi^\parallel_{\mu_5} \,,\qquad
 \langle J_\perp^2 \rangle_{\mu_5}
  = \langle J_\perp \rangle_{\mu_5}^2
  + \chi^\perp_{\mu_5} \,.
\label{eq:susceptibility}
\end{equation}
We note that $\langle\cdots\rangle_{\mu_5}$ and $\chi_{\mu_5}$ above
are an expectation value and susceptibility of the electric current,
respectively, for a fixed value of $\mu_5$.  Under our assumptions
listed above, the ensemble average reads,
\begin{align}
 &\llangle \cos(\Dphi_+ + \Dphi_+) \rrangle
  = -\llangle \cos(\Dphi_+ + \Dphi_-) \rrangle \notag\\
 &= \frac{c}{N_\pm^2} \int\rmd\mu_5\; \W(\mu_5) \bigl(
 \langle J_\perp^2 \rangle_{\mu_5}
 - \langle J_\parallel^2 \rangle_{\mu_5} \bigr) \notag\\
 &= -\frac{c}{N_\pm^2} \int\rmd\mu_5\; \W(\mu_5) \bigl(
 \langle J_\parallel \rangle_{\mu_5}^2 + \chi_{\mu_5}^\parallel
  - \chi_{\mu_5}^\perp \bigr) \,,
\label{eq:cosDphi}
\end{align}
where $\W(\mu_5)$ is the weight function normalized such that
$\int\rmd\mu_5\,\W(\mu_5)=1$ (and thus $\W(\mu_5)$ has a mass
dimension in our definition).  We will specify this function later.
The overall constant, $c$, is treated as a free parameter which is
independent of collision geometry, and we approximate
$N_+\approx N_-\approx N_\pm$.

It is interesting to see that the combination of the transverse and
longitudinal current correlators in Eq.~(\ref{eq:cosDphi}) has a clear
correspondence to the form of Eq.~(\ref{eq:voloshin}).  In
$\langle J_\perp^2\rangle_{\mu_5}-\langle J_\parallel^2\rangle_{\mu_5}$
together with Eq.~(\ref{eq:susceptibility}), there is a cancellation
between major components in $\chi_{\mu_5}^\perp$ and
$\chi_{\mu_5}^\parallel$, that is interpreted as the cancellation
between backgrounds $B^{\text{in}}_{\alpha\beta}$ and
$B^{\text{out}}_{\alpha\beta}$ discussed above.  However, because of
the presence of the external magnetic field the cancellation is not
necessarily exact, which gives a background on top of the CME
contribution from $\langle J_\parallel\rangle_{\mu_5}^2$.


\section{Current and Susceptibility}
\label{sec:susceptibility}
If we neglect the quantum fluctuation of gauge fields on top of the
external magnetic field $B$, we can explicitly accomplish the
integration in the quark sector with full inclusion of $B$.  We can
write down the thermodynamic grand potential at finite $T$ and
$\mu_q$, which is defined by $\Omega=-T\ln Z$ from the partition
function $Z$, given as~\cite{Fukushima:2008xe},
\begin{equation}
 \Omega = -V N_c \sum_f \frac{|q_f B|}{2\pi}\sum_{s=\pm}
  \sum_{k=0}^\infty \alpha^f_{ks} \int_{-\infty}^\infty \!
  \frac{\rmd p_z}{2\pi} \bigl[ \omega_{k\lambda}+T \sum_{\pm}
  \ln\bigl(1+\rme^{-(\omega_{k\lambda}\pm\mu_q)/T} \bigr)\bigr] \,,
\label{eq:Omega}
\end{equation}
where $f$, $s$, and $k$ refer to the flavor, spin, and Landau level
indices, respectively.  The factor $\alpha^f_{ks}$ is defined by
\begin{equation}
 \alpha^f_{ks} \equiv \left\{ \begin{array}{ll}
  \delta_{s+} & \text{ for~~~ $k=0$  ~~and~~ $q_f B>0$} \,,\\
  \delta_{s-} & \text{ for~~~ $k=0$ ~~and~~ $q_f B<0$} \,,\\
  1 & \text{ for~~~ $k>0$} \,,
 \end{array} \right.
\end{equation}
which takes care of the spin degeneracy depending on the zero or
non-zero modes.  The quasi-particle dispersion relation is
\begin{equation}
 \omega_{k\lambda} = \sqrt{ \Bigl(\sqrt{p_z^2 +
  2|q^fB|k} + \lambda \mu_5 \Bigr)^2 + M_f^2} \,.
\end{equation}
Here $\lambda\equiv \sgn(p_z)s$ stands for the helicity and $M_f$
represents the quark mass for flavor $f$.  We note that
$\omega_{k\lambda}$ is a flavor dependent quantity, though we omit an
index $f$ for concise notation.  In our convention $p_z$ refers to the
$z$-component of the three momentum $\bp$ (not the third component of
$p_\mu$ which has an additional minus sign from the metric).


\subsection{Differentiating the grand potential}

We can calculate the expectation value and the susceptibility of the
electric current operator by taking a functional derivative with
respect to the (source) gauge field $A_\mu$ as discussed in
Ref.~\cite{Fukushima:2008xe}.  In general the electric current can be
expressed as
\begin{equation}
 \langle j^\mu(x) \rangle = -\frac{\delta\Gamma[A]}{\delta A_\mu(x)}
  \biggr|_{A=\bar{A}} \,,
\end{equation}
where $\Gamma[A]$ is the effective action in a certain gauge and
$\bar{A}$ represents the background gauge field (corresponding to the
external magnetic field in our case).  In the finite-temperature field
theory the effective action (potential) translates into the
thermodynamic potential.  Thus, the induced current is,
\begin{equation}
 \langle J^\mu \rangle_{\mu_5}
  = \int\rmd^4 x \,\frac{\delta\, T\ln Z[A]}{\delta A_\mu(x)}
  \biggr|_{A=\bar{A}}
  = -\int\rmd^4 x\, \frac{\delta\, \Omega[A]}{\delta A_\mu(x)}
  \biggr|_{A=\bar{A}} \,.
\label{eq:current}
\end{equation}
It is easy to write down an expression for the current in
configuration space using the above equation with the functional
derivative, which turns out to be equivalent with the diagrammatic
method (see Appendix~\ref{app:curprop}).  Because
$\langle j^\mu(x)\rangle$ does not have $x$ dependence under a
spatially and temporally homogeneous magnetic field, we can replace
the functional derivative by a derivative with respect to homogeneous
$A_\mu$, that is,
\begin{equation}
 \langle J^\mu \rangle_{\mu_5}  =
  -\frac{\rmd \Omega[A]}{\rmd A_\mu} \biggr|_{A=\bar{A}} \,.
\label{eq:current2}
\end{equation}
The advantage of this rewriting is that we can now directly work in
momentum space.  Then, because of the structure of the covariant
derivative $\bp-q_f\bA$ (where $q_f>0$ for \textit{positively} charged
flavor), the derivative with respect to $A_\mu$ can be replaced by
that with respect by $p_\mu$ times $-q_f$.  To evaluate the current we
have to take the derivative with respect to $A_3$ which becomes the
derivative with respect to $-p_z$ times $-q_f$.  In this way the
current parallel to the external magnetic field $B$ is immediately
written down as~\cite{Fukushima:2008xe}
\begin{align}
 \langle J_\parallel \rangle_{\mu_5}
 &= V N_c \sum_{f,s,k} \frac{q_f|q_f B|}{2\pi}\alpha^f_{ks}
  \int_{-\infty}^\infty\! \frac{\rmd p_z}{2\pi}
  \frac{\rmd}{\rmd p_z} \bigl[ \omega_{k\lambda}+T \sum_{\pm}
  \ln\bigl(1+\rme^{-(\omega_{k\lambda}\pm\mu_q)/T} \bigr)\bigr]
  \notag\\
 &= V N_c\sum_{f,s} \frac{q_f|q_f B|}{2\pi} \alpha^f_{0s}\;
  \frac{1}{2\pi} \; 2s\mu_5
 = V N_c\sum_f \frac{q_f^2 B\mu_5}{2\pi^2} \,.
\label{eq:curmu5}
\end{align}
Here we have used the fact that the spin sum ($s=\pm$) makes a
cancellation for non-zero modes and only the zero-mode contribution
from $k=0$ remains non-vanishing.  The current has an origin in the
quantum anomaly coming from the surface term of the $p_z$-integration,
so it is an exact result and insensitive to any infrared scales such
as the temperature $T$, chemical potential $\mu_q$, and quark masses
$M_f$ (see also
Refs.~\cite{Alekseev:1998ds,Metlitski:2005pr,Newman:2005as}).  Since
rotational symmetry in the transverse plane perpendicular to the
magnetic field is kept unbroken, the transverse currents are zero;
$\langle J_\perp\rangle_{\mu_5}=0$.  What we have addressed so far is
just to remind the discussions given in our previous
work~\cite{Fukushima:2008xe}.  We present an alternative derivation of
the induced current using the exact propagator in a magnetic field in
Appendix~\ref{app:curprop}.

In contrast to the current expectation value, the (unrenormalized)
current susceptibility is dominated mostly by non-zero modes at high
Landau levels; this makes the susceptibility (but not the difference
of parallel and transverse susceptibilities) sensitive to the UV
regularization, making it unphysical.  Interestingly enough, the
zero-mode again plays an essential role in a finite difference of
$\chi^\parallel_{\mu_5}-\chi^\perp_{\mu_5}$.  The susceptibility is
deduced from
\begin{equation}
 \chi = -T\, \frac{\delta^2\, \Omega}{\delta A_\mu^2} \,,
\end{equation}
in the same way as the previous discussions on the induced current.
The parallel component is again easy to evaluate because the structure
of the covariant derivative is unchanged.  We must here introduce a UV
cutoff $\Lambda$.  We shall impose $\Lambda$ in a symmetric way, that
is, by Heaviside's step function
$\theta(\Lambda^2 \!-\! p_z^2 \!-\! 2|q_f B|k)$.  We note that our
final result of $\chi_{\mu_5}^\parallel-\chi_{\mu_5}^\perp$ is UV
finite and thus it does not depend on the cutoff parameter nor
scheme.  Our naive way to cut the momentum integration off is not
gauge invariant, and so it is of no use in order to compute
$\chi_{\mu_5}^\parallel$ and $\chi_{\mu_5}^\perp$ individually.

In our prescription the $p_z$-integration is bounded by
$\Lambda_k\equiv \sqrt{\Lambda^2-2|q_f B|k}$ for a given $k$, where
$k$ takes a value from $0$ to
$k_\Lambda\equiv\lfloor\Lambda^2/(2|q_f B|)\rfloor$.
Thus, by differentiating the grand potential with respect to $A_z$
twice, we have the longitudinal current susceptibility as
\begin{align}
 \chi^\parallel_{\mu_5}
 &= VT N_c \sum_{f,s,k} \frac{q_f^2|q_f B|}{2\pi}\alpha^f_{ks}
  \int_{-\Lambda_k}^{\Lambda_k}\! \frac{\rmd p_z}{2\pi}
  \frac{\rmd^2}{\rmd p_z^2} \bigl[ \omega_{k\lambda}+T \sum_{\pm}
  \ln\bigl(1+\rme^{-(\omega_{k\lambda}\pm\mu_q)/T} \bigr)\bigr] \notag\\
 &= VT N_c \sum_{f,s,k} \frac{q_f^2|q_f B|}{4\pi^2}
  \alpha^f_{ks}\frac{p_z}{\omega_{k\lambda}}\biggl( 1 \!+\!
  \frac{\lambda\mu_5}{\sqrt{p_z^2+2|q_f B|k}} \biggr)
  \bigl[ 1\!-\! n_{_F}(\omega_{k\lambda}) \!-\!
  \bar{n}_{_F}(\omega_{k\lambda})
  \bigr] \biggr|^{p_z=\Lambda_k}_{p_z=-\Lambda_k} \notag\\
 &= VT N_c \sum_{f,s,k} \frac{q_f^2|q_f B| g_k}{4\pi^2}
  \frac{\Lambda_k}{\omega_{\Lambda \lambda}^f}\Bigl(1 +
  \frac{s\mu_5}{\Lambda} \Bigr) \bigl[ 1-n_{_F}(\omega_{\Lambda s})
  -\bar{n}_{_F}(\omega_{\Lambda s}) \bigr] \,,
\label{eq:chi_parallel}
\end{align}
where $\omega_{\Lambda s}\equiv \sqrt{(\Lambda+s\mu_5)^2+M_f^2}$ and
the Fermi-Dirac distribution functions are
$n_{_F}(\omega)\equiv[\rme^{(\omega-\mu_q)/T}+1]^{-1}$ and
$\bar{n}_{_F}(\omega)\equiv[\rme^{(\omega+\mu_q)/T}+1]^{-1}$.  We have
introduced a new notation $g_k$ which is the spin degeneracy defined
by
\begin{equation}
 g_k \equiv \alpha_{ks}^f + \alpha_{k-s}^f
  = \left\{ \begin{array}{lp{3mm}l}
  1 && \text{for $k=0$} \\ 2 && \text{for $k\neq0$}
  \end{array} \right. \,.
\end{equation}
In principle, the current susceptibility has an additional
contribution from mixing with the chiral susceptibility $\chi_M$
through dependence of the dynamical mass $M_f$ on $A_\mu$, resulting
in a contribution like $(\rmd M_f/\rmd A_\mu)^2 \chi_M$.  This is,
however, negligible for the strength of magnetic fields and associated
currents relevant to heavy-ion collisions.  (We have numerically
confirmed this by using a mean-field chiral model.)

The longitudinal susceptibility~(\ref{eq:chi_parallel}) is UV
divergent, as we have already mentioned, and is strongly dependent on
the value of cut-off $\Lambda$ and how $\Lambda$ is imposed.  In fact
it is straightforward to take the limit of $B=\mu_5=0$ in
Eq.~(\ref{eq:chi_parallel}), which leads to an orientation independent
susceptibility;
\begin{equation}
 \chi_0 = VT N_c \sum_f \frac{q_f^2 \Lambda^3}{3\pi^2
  \omega_\Lambda} \bigl[ 1-n_{_F}(\omega_\Lambda)
  -\bar{n}_{_F}(\omega_\Lambda) \bigr] \,,
\label{eq:chi0}
\end{equation}
where we have defined $\omega_\Lambda\equiv \sqrt{\Lambda^2+M_f^2}$.
This is UV divergent proportional to $\Lambda^2$ for large $\Lambda$.
It is necessary, therefore, to perform the renormalization to extract
a finite answer for $\chi^\parallel_{\mu_5}-\chi_0$, in which the
leading divergence $\sim\Lambda^2$ cancels but the logarithmic
divergence may remain in general.  As long as
$\chi_{\mu_5}^\parallel-\chi_{\mu_5}^\perp$ is concerned, as we are
explicitly computing, only a finite term appears and there is no
subtlety associated with UV divergence and renormalization.

Evaluating $\chi^\perp_{\mu_5}$ is not as easy as
$\chi^\parallel_{\mu_5}$;  we cannot simply replace $\bp$ by one
augmented with $\bA$ in the transverse direction because of the Landau
quantization.  In what follows below we will explain how to calculate
this quantity by taking three steps.

\noindent
\textbf{Step i})~~
We should return to the original definition of the current correlation
function in terms of quark fields;
$\chi_i\propto \int\!\rmd^4 x\rmd^4y\langle\bar{\psi}\gamma^i\psi(x)
\bar{\psi}\gamma^i\psi(y)\rangle$.  We already know the answer when
$B=\mu_5=0$, which is given in Eq.~(\ref{eq:chi0}).  In the first
step, hence, we will confirm this known answer by the diagrammatic
method.

\noindent
\textbf{Step ii)}~~
We will introduce an external magnetic field $B$ in the second step.
The quark propagator requires a modification with the projection
operator in Dirac indices.  Then we will find that the transverse
susceptibility takes the same form as the longitudinal one except for
the Landau zero-mode.

\noindent
\textbf{Step iii)}~~
Finally we will extend the calculation to the case with finite
$\mu_5$.  To compute the quark propagator we need to insert one more
projection operator with respect to the Dirac indices, which separates
opposite helicity states.


\subsection{Establishing the diagrammatic method --- step i)}

In Minkowskian convention we can express the electric-current
correlation function as
\begin{equation}
 \chi_i = \rmi\, T^2 N_c\sum_f q_f^2 \,
  \int\rmd^4x\,\rmd^4y\,
  \bigl\langle \bar{\psi}(x)\gamma^i \psi(x) \,
  \bar{\psi}(y) \gamma^i \psi(y)\bigr\rangle \,.
\label{eq:chi_def}
\end{equation}
By taking the contraction of quark fields, the above expression is
decomposed into the disconnected (tadpole-type) and connected
contributions.


\begin{figure}[ht]
\begin{center}
 \includegraphics[scale=0.8]{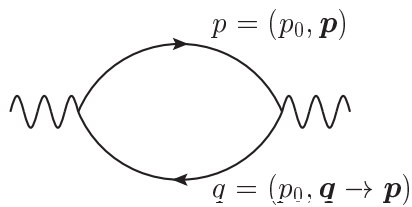}
 \caption{Diagram of the connected contribution to the
   electric-current correlation function (photon self-energy) and the
   momenta running in the loop.}
 \label{fig:diagram}
\end{center}
\end{figure}


The diagrammatic representation of the connected contribution from
Eq.~(\ref{eq:chi_def}) is displayed in Fig.~\ref{fig:diagram}, from
which we can recognize that this is nothing but the diagram for the
photon self-energy (polarization tensor, see Ref.~\cite{Tsai:1974ap}
for discussions on the vacuum polarization in a magnetic field).
Therefore the answer must be zero in the limit of zero momentum
insertion from the external legs;  otherwise the photon becomes
massive and breaks gauge invariance.  It is obvious from such a
recognition that the UV divergent answer in Eq.~(\ref{eq:chi0}) is
unphysical and merely an artifact from the naive cutoff prescription.

The corresponding expression to Fig.~\ref{fig:diagram} translates into
the current susceptibility of our interest here given in the following
form;
\begin{equation}
 \chi_i = -\rmi\, T^2 N_c\sum_f q_f^2 \,
  \int\rmd^4x\,\rmd^4y\,\tr
  \bigl[\gamma^i G(x,y) \gamma^i G(y,x)\bigr] \,,
\label{eq:chi_general}
\end{equation}
where $G(x,y)$ represents the quark propagator in configuration
space.  Integrating over $x$ and $y$ and using the familiar expression
of the propagator in momentum space for $B=\mu_5=0$ (then there is no
preferred direction and so we put a subscript $0$ instead of $i$) we
immediately reach
\begin{align}
 \chi_0 &= -\rmi\, VT N_c\sum_f q_f^2
  \int\frac{\rmd^3 p}{(2\pi)^3}
  \int^T \!\frac{\rmd p_0}{2\pi}\, \tr\biggl[ \gamma^i \;
  \frac{\rmi}{\feyn{q}-M_f} \;\gamma^i\;
  \frac{\rmi}{\feyn{p}-M_f} \biggr] \notag\\
 &= \rmi\, VT N_c\sum_f q_f^2 \int\frac{\rmd^3 p}{(2\pi)^3}
  \int^T \!\frac{\rmd p_0}{2\pi}\,
  \frac{4(2p_i q_i +p\!\cdot\! q -M_f^2)}
  {(p_0^2-\omega_q^2)(p_0^2-\omega_p^2)} \,,
\label{eq:chi0_diagram}
\end{align}
where $p=(p_0=2\pi\rmi(n+\half)T+\mu_q,\bp)$ and $q=(p_0,\bq\to\bp)$
and $\int^T(\rmd p_0/2\pi) \equiv \rmi\,T\sum_n$.  We can carry out
the Matsubara frequency sum over $n$ to find,
\begin{align}
 & \rmi \int^T\!\frac{\rmd p_0}{2\pi}
  \frac{4(2p_i q_i +p\!\cdot\! q-M_f^2)}
  {(p_0^2-\omega_q^2)(p_0^2-\omega_p^2)} \notag\\
 & = 4\Biggl\{ \frac{1}{2(\omega_p + \omega_q)}
  \Biggl[ 1 - \frac{\omega_q \bigl( n_{_F}(\omega_q)\!+\!
  \bar{n}_{_F}(\omega_q) \bigr) \!-\! \omega_p \bigl(
  n_{_F}(\omega_p)\!+\!\bar{n}_{_F}(\omega_q) \bigr)}
  {\omega_q - \omega_p} \Biggr] \notag\\
 &\quad -\frac{(2p_i q_i \!-\! \bp \!\cdot\! \bq \!-\! M_f^2)
   \bigl[\omega_q^{-1} \bigl( n_{_F}(\omega_q)\!+\!
   \bar{n}_{_F}(\omega_q) \bigr)
  -\omega_p^{-1} \bigl( n_{_F}(\omega_p)\!+\!
  \bar{n}_{_F}(\omega_p)\bigr)\bigr]}
  {2(\omega_p + \omega_q)(\omega_q-\omega_p)} \Biggr\} \notag\\
 & \underset{\boldsymbol{q}\to\boldsymbol{p}}{\longrightarrow}
  2\Biggl\{ \biggl( \frac{1}{\omega_p}-\frac{p_i^2}{\omega_p^3}
  \biggr)\bigl[ 1-n_{_F}(\omega_p)-\bar{n}_{_F}(\omega_p) \bigr]
  + \frac{p_i^2}{\omega_p^2}
  \frac{\rmd \bigl[ 1-n_{_F}(\omega_p)-\bar{n}_{_F}(\omega_p)
  \bigr]} {\rmd\, \omega_p} \Biggr\} \notag\\
 &= 2\frac{\rmd}{\rmd p_i} \biggl\{ \frac{p_i}{\omega_p}
  \bigl[ 1-n_{_F}(\omega_p)-\bar{n}_{_F}(\omega_p) \bigr]
  \biggr\} \,.
\label{eq:sum_n}
\end{align}
Plugging the above (\ref{eq:sum_n}) into Eq.~(\ref{eq:chi0_diagram})
and integrating over the momentum $p$ with a UV cutoff $\Lambda$, we
can correctly rederive the result of Eq.~(\ref{eq:chi0}) in the
diagrammatic way.

Before closing this subsection we should mention on our procedure to
take the limit of $\bq\to\bp$.  In this subsection this procedure is
not indispensable;  we could have started the calculation with
$\bq=\bp$.  Then, the residue of a double pole gives a term involving
the derivative of the distribution functions with respect to
$\omega_p$ which is exactly the same as the term arising from the
$\bq\to\bp$ limit.  The reason why we distinguished $q$ from $p$ is
that, as we will see shortly, the presence of $B\neq0$ induces a
difference in the transverse component in Landau-quantized momenta.
Thus, the calculation process elucidated in this subsection has a
smooth connection to the later generalization.


\subsection{Turning on $B$ --- step ii)}

In the presence of non-zero $B$ the transverse and longitudinal
directions become distinct even before introducing finite $\mu_5$,
which should result in $\chi_0^\parallel-\chi_0^\perp\neq0$ regardless
of topological excitations.  We already know the answer for the
longitudinal susceptibility from Eq.~(\ref{eq:chi_parallel}), that is,
by taking the limit of $\mu_5=0$, we can readily write down,
\begin{equation}
 \chi_0^\parallel = VT N_c \sum_{f,k}
  \frac{q_f^2|q_f B|g_k}{2\pi^2}
  \frac{\Lambda_k}{\omega_\Lambda}
  \bigl[ 1-n_{_F}(\omega_\Lambda)
  -\bar{n}_{_F}(\omega_\Lambda) \bigr] \,.
\label{eq:chi0_para}
\end{equation}
Let us calculate the transverse susceptibility by means of the
diagrammatic method developed in the previous subsection.  For this
purpose we should first come by the quark propagator under a constant
magnetic field.

The solution of the Dirac equation with a constant magnetic field is
known, which forms the complete set of orthogonal wave-functions.  We
here introduce several new notations.  We use the following basis
functions in a gauge choice where $A_0=A_x=A_z=0$ and $A_y=Bx$;
\begin{equation}
 \begin{split}
 & f_{k+}(x) = \phi_k(x-p_y/(qB))  \qquad (k=0,1,2,\dots) \,, \\
 & f_{k-}(x) = \phi_{k-1}(x-p_y/(qB))  \qquad (k=1,2,3,\dots) \,.
 \end{split}
\label{eq:f}
\end{equation}
with $\phi_k(x)$ being the standard Landau-quantized wave-function
defined by
\begin{equation}
 \phi_k(x) = \sqrt{\frac{1}{2^k k!}} \biggl(\frac{|qB|}{\pi}
  \biggr)^{1/4} \!\exp\Bigl(-\frac{1}{2}|qB|x^2\Bigr)
  H_k\bigl(\sqrt{|qB|}x\bigr) \,,
\end{equation}
where $H_k(x)$ represents the Hermite polynomial of degree $k$.  (We
omit the subscript $f$ of $q_f$ for the moment.)  We can easily
confirm that the basis functions satisfy the orthogonality property as
follows;
\begin{equation}
 \begin{split}
 & \int \rmd x\, f_{k+}(x) f_{l+}(x) = \delta_{k,l} \,,\\
 & \int \rmd x\, f_{k+}(x) f_{l-}(x) = \delta_{k,l-1}
  \quad (l\ge1) \,, \\
 & \int \rmd x\, f_{k-}(x) f_{l-}(x) = \delta_{k,l} \,,\\
 & \int \rmd x\, f_{k-}(x) f_{l+}(x) = \delta_{k-1,l}
  \quad (k\ge1) \,.
 \end{split}
\label{eq:orthogonal}
\end{equation}
We can define the projection matrix with respect to the Dirac index
according to Ritus' method~\cite{Ritus:1972ky} as
\begin{equation}
 P_k(x) = \frac{1}{2}\bigl[ f_{k+}(x) + f_{k-}(x) \bigr]
  +\frac{\rmi}{2}\bigl[ f_{k+}(x) - f_{k-}(x) \bigr]\,
  \gamma^1 \gamma^2 \,,
\label{eq:Pn}
\end{equation}
for $qB>0$ and $f_{k+}$ and $f_{k-}$ are swapped for the case with
$qB<0$.  It is then straightforward to prove that
\begin{equation}
 \begin{split}
 &\bigl( \rmi \feyn{\partial} - q\feyn{A} - M_f \bigr)P_k(x)
  \,\rme^{-\rmi(p_0 t-p_y y-p_z z)} \\
 &\qquad
 = P_k(x) \bigl( p_0\gamma^0 + \sgn(qB)\sqrt{2|qB|k}\gamma^2
  -p_z\gamma^3 - M_f \bigr)\,
  \rme^{-\rmi(p_0 t-p_y y-p_z z)} \,,
 \end{split}
\end{equation}
where the right-hand side is expressed in a form of the free Dirac
operator with a modified momentum
$\tilde{p}=(p_0,0,-\sgn(qB)\sqrt{2|qB|k},p_z)$.  Hence,
the solution of the Dirac equation with a magnetic field is given by a
combination of the projection matrix and the free Dirac spinors
$u(p,s)$ for particles and $v(p,s)$ for antiparticles with a momentum
argument $\tilde{p}$, that is,
\begin{equation}
 P_k(x)\, u(\tilde{p},s)\,
 \rme^{-\rmi(p_0 t-p_y y-p_z z)} \,, \qquad
 P_k(x)\, v(\tilde{p},s)\,
 \rme^{\rmi(p_0 t-p_y y-p_z z)} \,.
\end{equation}
Here we note that $P_k(x)$ is a real function.  Using these
complete-set functions and defining associated creation and
annihilation operators, we can compute the propagator
$\langle \psi(x)\bar{\psi}(y)\rangle$ to reach
\begin{equation}
 \begin{split}
 G(x,y) &= \langle \psi(x)\bar{\psi}(y) \rangle \\
 & = \int\frac{\rmd p_0}{2\pi} \int
  \frac{\rmd p_y\,\rmd p_z}{(2\pi)^2} \sum_k \,
  \rme^{-\rmi[p_0(x_0-y_0)-p_y(x_y-y_y)-p_z(x_z-y_z)]} \\
 &\qquad\qquad\qquad\qquad\qquad\quad\times
  P_k(x)\,\rmi\, (\feyn{\tilde{p}} - M_f)^{-1} P_k (y) \,.
 \end{split}
\label{eq:propagator}
\end{equation}
It should be noted that $P_k(x)$ and $P_k(y)$ have implicit $p_y$
dependence in the argument (see Eq.~(\ref{eq:f})).  The spin
degeneracy is automatically taken into account by the property that
$f_{0-}(x)=0$ and thus only one spin state has a non-zero contribution
for $k=0$.

\begin{figure}[h]
\begin{center}
 \includegraphics[scale=0.8]{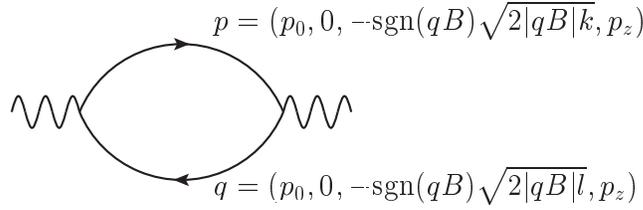}
 \caption{Diagram and the running momenta corresponding to
   Eq.~(\ref{eq:chi_i_b}).  The Landau levels denoted by $k$ and $l$
   are constrained (either $k=l$ or $k=l+1$ or $l=k+1$) by the
   projection operators.}
 \label{fig:diagram2}
\end{center}
\end{figure}

We use this form of the quark propagator to express the
susceptibility~(\ref{eq:chi_general}).  Then we have to perform the
coordinate integration with respect to $x_0$, $y_0$, $x_x$, $y_x$,
$x_y$, $y_y$, $x_z$, $y_z$, and the momentum integration over $p_0$,
$p_0'$, $p_y$, $p_y'$, $p_z$, $p_z'$ and also the summation over $k$
and $l$, where $p_0'$, $p_y'$, $p_z'$, and $l$ are from another
$G(y,x)$.  As usual, after we integrate with respect to $x_0$, $x_y$,
and $x_z$, we have
$(2\pi)^3\delta(p_0-p_0')\delta(p_y-p_y')\delta(p_z-p_z')$ (there is
no momentum insertion from external legs), from which only the
integrations with respect to $p_0$, $p_y$, and $p_z$ remain.  Then,
renaming $x_x\to x$ and $y_x\to y$, we have
\begin{equation}
 \begin{split}
 \chi_i &= \rmi L_y L_z T N_c\sum_f q_f^2 \int\frac{\rmd p_y}{2\pi}
  \sum_{k,l}
  \int\frac{\rmd p_z}{2\pi} \int^T\!\frac{\rmd p_0}{2\pi}
  \int \rmd x\, \rmd y \\
 &\qquad\qquad \times \tr\Bigl[ \gamma^i P_k(x)
  (\feyn{\tilde{p}} - M_f)^{-1} P_k(y) \gamma^i
  P_l(y)(\feyn{\tilde{q}} - M_f)^{-1} P_l(x) \Bigr] \,,
 \end{split}
\end{equation}
where we have defined
$\tilde{q}=(p_0,0,-\sgn(qB)\sqrt{2|qB|l\,},p_z)$,
$L_y\equiv\int\rmd y_y$, and $L_z\equiv\int\rmd y_z$.  Because $p_y$
is now common in all $P_k(x)$, $P_l(y)$, etc, a shift in the
integration variables, $x$ and $y$, can get rid of $p_y$ from the
integrand.  Therefore, $\int(\rmd p_y/2\pi)$ simply counts the number
of Landau-quantized states, that is given by $L_x (|qB|/2\pi)$.  Here
$L_x L_y L_x$ is nothing but the volume $V$.  After all we can write
the above into a form of
\begin{equation}
 \begin{split}
 \chi_i &= \rmi\, VT N_c\sum_f \frac{q_f^2|q_f B|}{2\pi}
  \sum_{k,l}
  \int\frac{\rmd p_z}{2\pi} \int^T\!\frac{\rmd p_0}{2\pi}
  \int \rmd x\, \rmd y \\
 &\qquad\qquad \times \tr\Bigl[ \gamma^i P_k(x)
  (\feyn{\tilde{p}} - M_f)^{-1} P_k(y) \gamma^i
  P_l(y)(\feyn{\tilde{q}} - M_f)^{-1} P_l(x) \Bigr] \,.
 \end{split}
\label{eq:chi_i_b}
\end{equation}
Apart from the projection operator which we will discuss soon below,
this expression has a diagrammatic representation with momenta running
in the loop as depicted in Fig.~\ref{fig:diagram2}.


Because $P_k(x)$ is a combination of the unit matrix and
$\gamma^1\gamma^2$ as defined in Eq.~(\ref{eq:Pn}) and thus is
commutative with $\gamma^3$, the longitudinal susceptibility is easy
to evaluate.  Using
$P_l(x)\gamma^3 P_k(x)=P_l(x)P_k(x)\gamma^3$ and the
orthogonality~(\ref{eq:orthogonal}), after integration over $x$, we
obtain $\delta_{l,k}$ for $k>0$ and
$(1+\rmi\,\sgn(qB)\gamma^1\gamma^2)/2$ for $k=l=0$.  The calculation
is then reduced to the same as in the previous case with several
replacements;  the momentum $p$ by $\tilde{p}$, and the integration
with respect to transverse momenta by the summation over the Landau
levels.  Then the longitudinal result of Eq.~(\ref{eq:chi0_para}) is
almost trivially understood from Eq.~(\ref{eq:sum_n}) with $i=z$.

We are now ready to consider the transverse polarization.  In this
case $\gamma^1$ and $\gamma^2$ are not commutative with $P_k(x)$.
Because the calculations for $\gamma^1$ and $\gamma^2$ are just
parallel, we focus only on the case of $\gamma^1$ below.  Using
\begin{equation}
 \begin{split}
 & \int\rmd x\, P_l(x)\gamma^1 P_k(x) \\
 &\quad =
  \left\{ \begin{array}{lp{3mm}l}
   \displaystyle
   \frac{1}{2}\delta_{l-1,k}\gamma^1
   -\frac{\rmi}{2}\delta_{l-1,k}\gamma^2 && \text{for $k=0$} 
   \vspace{0.5em}\\
   \displaystyle
   \frac{1}{2}\delta_{l,k-1}\gamma^1
   +\frac{\rmi}{2}\delta_{l,k-1}\gamma^2 && \text{for $l=0$} 
   \vspace{0.5em}\\
   \displaystyle
   \frac{1}{2}\bigl[ \delta_{l-1,k}+\delta_{l,k-1} \bigr]\gamma^1
   -\frac{\rmi}{2}\bigl[ \delta_{l-1,k}-\delta_{l,k-1} \bigr]
   \gamma^2 && \text{for $k\neq0$, $l\neq0$}
  \end{array} \right. \,.
 \end{split}
\label{eq:delta_trans}
\end{equation}
we can evaluate the trace with respect to the Dirac index to arrive
finally at
\begin{equation}
 \chi_0^\perp = \rmi\,VT N_c\sum_f   \frac{q_f^2|q^fB|}{2\pi}
  \sum_{k,l}
  \int\frac{\rmd p_z}{2\pi}
  \int^T\!\frac{\rmd p_0}{2\pi}\,
  \frac{4(p_0^2-p_z^2-M_f^2)\,\delta_{l-1,k}}
  {(p_0^2-\omega_q^2)(p_0^2-\omega_p^2)} \,.
\end{equation}
It should be mentioned that there emerges no spin degeneracy factor
in the above expression.  Taking the Matsubara sum leads to
\begin{align}
 & \rmi \int^T\!\frac{\rmd p_0}{2\pi} \sum_l \,
  \frac{2(p_0^2-p_z^2-M_f^2)\,\delta_{l-1,k}}
  {(p_0^2-\omega_q^2)(p_0^2-\omega_p^2)} \notag\\
 &= \frac{k+1}{\omega_q}
  \bigl[ 1-n_{_F}(\omega_q)-\bar{n}_{_F}(\omega_q) \bigr]
  -\frac{k}{\omega_p} \bigl[ 1-n_{_F}(\omega_p)
  -\bar{n}_{_F}(\omega_p) \bigr]
\end{align}
with, as indicated in Fig.~\ref{fig:diagram2},
$p=(p_0=2\pi\rmi(n+\half)T+\mu_q,0,-\sgn(qB)\sqrt{2|qB|k},p_z)$ and
$q=(p_0,0,-\sgn(qB)\sqrt{2|qB|l\,},p_z)$, where $l=k+1$.  Here we note
that $k$ and $l$ are both bounded by the UV cutoff, that is, the
maximum $k$ is $\bar{k}-1$ with
$\bar{k}\equiv \lfloor(\Lambda^2-p_z^2)/(2|q_f B|)\rfloor$, and $p_z$
is cut off by $\Lambda$.  We realize then that subsequent terms are
canceled in summation over $k$ and only the edge term (from
$k=\bar{k}-1$) remains non-vanishing, which is reminiscent of the
remaining surface term in the $B=0$ case out of the momentum
integration of Eq.~(\ref{eq:sum_n}).  Therefore, at last, we get
\begin{equation}
 \chi_0^\perp = VT N_c\sum_f \frac{q_f^2|q_f B|}{2\pi^2}
  \int_{-\Lambda}^\Lambda \rmd p_z \,
  \frac{\bar{k}}{\omega_\Lambda}
  \bigl[ 1-n_{_F}(\omega_\Lambda)-\bar{n}_{_F}(\omega_\Lambda)
  \bigr]
\label{eq:chi0_perp}
\end{equation}
after the summation with respect to $k$ from $0$ to $\bar{k}-1$.  To
be more precise, it should be mentioned here that we approximated as
$2|q_f B|\bar{k}+p_z^2 + M_f^2 \to \omega_\Lambda^2 = \Lambda^2 + M_f^2$,
which is valid as long as $\Lambda^2 \gg |q_f B|$.

It is very interesting that we can find a simple analytical formula
for the difference $\chi_0^\parallel-\chi_0^\perp$.  From
Eqs.~(\ref{eq:chi0_para}) and (\ref{eq:chi0_perp}) we can write,
\begin{equation}
 \begin{split}
 & \chi_0^\parallel - \chi_0^\perp \\
 & = VT N_c \sum_f
  \frac{q_f^2|q_f B|}{2\pi^2} \frac{1}{\omega_\Lambda}
  \bigl[ 1\!-\!n_{_F}(\omega_\Lambda)\!-\!\bar{n}_{_F}(\omega_\Lambda) \bigr]
  \biggl( \sum_{k=0} g_k\Lambda_k - 2\int_0^\Lambda
  \rmd p_z \, \bar{k} \biggr) \,,
 \end{split}
\end{equation}
where, as we have already defined,
$\Lambda_k=\sqrt{\Lambda^2-2|q_f B|k}$ and
$\bar{k}=\lfloor(\Lambda^2-p_z^2)/(2|q_f B|)\rfloor$.  The point is
that we can rewrite the latter term in the parentheses into a form of
the discrete summation from the fact that $\bar{k}$ is defined by the
floor function.  That is,
\begin{equation}
 \begin{split}
 & \int_0^\Lambda \! \rmd p_z \, \bar{k}
  = \int_{\Lambda_1}^\Lambda \!\rmd p_z \times 0 +
  \int_{\Lambda_2}^{\Lambda_1} \!\rmd p_z \times 1 +
  \int_{\Lambda_3}^{\Lambda_2} \!\rmd p_z \times 2 + \cdots \\
 &= ( \Lambda_1 - \Lambda_2 )
  + 2( \Lambda_2 - \Lambda_3 )
  + 3( \Lambda_3 - \Lambda_4 )
  + \cdots = \sum_{k=1} \Lambda_k \,.
 \end{split}
\end{equation}
This form (times two) exactly looks like the first term in the
parentheses except for the $k=0$ term.  Therefore, only the zero-mode
with $k=0$ contributes to the final expression;
\begin{equation}
 \begin{split}
 \chi_0^\parallel - \chi_0^\perp &= VT N_c\sum_f
  \frac{q_f^2|q_fB|}{2\pi^2}\frac{\Lambda}{\omega_\Lambda}
  \bigl[ 1-n_{_F}(\omega_\Lambda)-\bar{n}_{_F}(\omega_\Lambda)
  \bigr] \\
 &\underset{\Lambda\to\infty}{\longrightarrow}
  VT N_c\sum_f \frac{q_f^2|q_f B|}{2\pi^2} \,.
 \end{split}
\label{eq:result_without_mu5}
\end{equation}
Although $\chi_0^\parallel$ and $\chi_0^\perp$ are divergent
$\sim \Lambda^2$ in the gauge-variant cutoff scheme, as we have
mentioned before, the difference between them is certainly UV finite,
so that it can be a well-defined quantity.


\subsection{Introducing $\mu_5$  --- step iii)}

To treat the situation in the presence of not only $B$ but also
$\mu_5$, we need to insert the projection matrices defined as
\begin{equation}
 \Gamma_\pm(p) \equiv \frac{1}{2} ( 1\pm \hat{\bp}\!\cdot\!
  \boldsymbol{\gamma} \gamma^0\gamma^5 ) \,,
 \label{eq:gammaprojector}
\end{equation}
to calculate the quark propagator which involves the inversion of
Dirac matrices.  The above-defined projection matrices have the
following property,
\begin{align}
 &\quad (\feyn{p}+\mu_5 \gamma^0\gamma^5 - M)^{-1}
  \Gamma_\pm(p) \notag\\
 &= (\feyn{p}+\mu_5 \gamma^0\gamma^5 + M)
  ( p^2-2\mu_5\bp\!\cdot\!\boldsymbol{\gamma}\gamma^0
  \gamma^5 -\mu_5^2-M^2)^{-1} \Gamma_\pm(p) \notag\\
 &= (\feyn{p}+\mu_5\gamma^0\gamma^5+M)
  \bigl[ p_0^2 - (|\bp|\pm\mu_5)^2 -M^2 \bigr]^{-1}
  \Gamma_\pm(p) \,.
\label{eq:diracoperatorinversion}
\end{align}
We can identify $\pm$ as $\lambda$ appearing in Eq.~(\ref{eq:Omega}),
so that $\Gamma_\pm(p)$ is the projection operator to the state with
helicity $\lambda=\pm$.  We insert the unity,
$1=\Gamma_+(\tilde{p})+\Gamma_-(\tilde{p})$, after the propagator to
take the inversion.  By doing this we can express the susceptibility
in the following way;
\begin{equation}
 \begin{split}
 \chi_i &= \rmi\, VT N_c\sum_f \frac{q_f^2|q_f B|}{2\pi}
  \sum_{k,l}
  \int\frac{\rmd p_z}{2\pi} \int^T\!\frac{\rmd p_0}{2\pi} \int\rmd x
  \,\rmd y \\
 &\qquad\times \tr\biggl[ \gamma^i P_k(x)
  (\feyn{\tilde{p}}+\mu_5\gamma^0\gamma^5 - M_f)^{-1}
  (\Gamma_+(\tilde{p})+\Gamma_-(\tilde{p})) P_k(y) \\
 &\qquad\qquad\times
  \gamma^i P_l(y)(\feyn{\tilde{q}} +\mu_5\gamma^0\gamma^5- M_f)^{-1}
  (\Gamma_+(\tilde{q})+\Gamma_-(\tilde{q}))P_l(x)
  \biggr] \,.
 \end{split}
 \label{eq:suscepmu5}
\end{equation}
Since we know the answer for the longitudinal susceptibility, let us
concentrate in calculating the transverse one (i.e.\ $i=x$)
hereafter.  After the integration over $x$ and $y$ with
Eq.~(\ref{eq:delta_trans}) we can find
\begin{equation}
 \begin{split}
 \chi_{\mu_5}^\perp &= \rmi\,VT N_c\sum_f \frac{q_f^2|q_f B|}
  {2\pi} \sum_{k,l} \int\frac{\rmd p_z}{2\pi}
  \int^T\!\frac{\rmd p_0}{2\pi} \\
 &\qquad\qquad \times \sum_{\lambda,\lambda'=\pm}
  \frac{\delta_{l-1,k}}{2}\,
  \frac{T^{11}_{\lambda\lambda'}+T^{22}_{\lambda\lambda'}
  +\rmi T^{12}_{\lambda\lambda'} -\rmi T^{21}_{\lambda\lambda'}}
  {(p_0^2-\omega_{p\lambda}^2 )(p_0^2-\omega_{q\lambda'}^2 )} \,.
 \end{split}
\end{equation}
where
\begin{equation}
 T^{ij}_{\lambda\lambda'} \equiv \tr\Bigl[
  \gamma^i (\feyn{\tilde{p}}+\mu_5\gamma^0\gamma^5 + M_f)
  \Gamma_\lambda(\tilde{p}) \gamma^j
  (\feyn{\tilde{q}}+\mu_5\gamma^0\gamma^5 + M_f)
  \Gamma_{\lambda'}(\tilde{q}) \Bigr] \,.
\end{equation}
We note that $\tr[\gamma^\mu\gamma^\nu\gamma^\rho\gamma^\sigma
\gamma^5]=-4\rmi\epsilon^{\mu\nu\rho\sigma}$ in the convention we
are using (where $\epsilon^{0123}=+1$).  After some lengthy
calculations we can notice that $T^{ij}_{\lambda\lambda'}$ becomes as
simple as
\begin{equation}
 \begin{split}
 T^{ij}_{\lambda\lambda'}
 & =   \bigl\{ \delta^{ij} +
   \lambda\lambda' \bigl[ (\hat{p}^i\hat{q}^j+\hat{p}^j\hat{q}^i)
   -\delta^{ij}\hat{\bp}\cdot\hat{\bq} \bigr]
   +\rmi\epsilon^{0ijk}(\lambda\hat{p}^k -\lambda' \hat{q}^k)
   \bigr\} \\
 &\quad\times \bigl[ p_0^2 + \lambda\lambda'
  (|\tilde{\bp}|+\lambda\mu_5)
  (|\tilde{\bq}|+\lambda'\mu_5) -M_f^2 \bigl] \,,
 \end{split}
\end{equation}
from which we get
\begin{equation}
 \begin{split}
 & T^{11}_{\lambda\lambda'}+T^{22}_{\lambda\lambda'}
  +\rmi T^{12}_{\lambda\lambda'} -\rmi T^{21}_{\lambda\lambda'} \\
 &= 2\bigl[p_0^2+\lambda\lambda'
  (|\tilde{\bp}|+\lambda\mu_5)(|\tilde{\bq}|+\lambda'\mu_5)
  -M_f^2 \bigr](1-\lambda\hat{p}^3)(1+\lambda'\hat{q}^3) \,.
 \end{split}
\end{equation}
The Matsubara sum amounts to
\begin{align}
 & \rmi \sum_{\lambda,\lambda'=\pm}\int^T\!\frac{\rmd p_0}{2\pi} \;
  \frac{\delta_{l-1,k}}{2}
  \frac{T^{11}_{\lambda\lambda'}+T^{22}_{\lambda\lambda'}
  -\rmi T^{12}_{\lambda\lambda'} +\rmi T^{21}_{\lambda\lambda'}}
  {(p_0^2-\omega_{p\lambda}^2)(p_0^2-\omega_{q\lambda'}^2)} \notag\\
 & = \sum_{\lambda,\lambda'=\pm}
  \frac{\delta_{l-1,k}}
  {2\omega_{q\lambda'}(\omega_{q\lambda'}^2-\omega_{p\lambda}^2)}
  \bigl[(|\tilde{\bq}|+\lambda'\mu_5)^2
  +\lambda\lambda'(|\tilde{\bp}|+\lambda\mu_5)
  (|\tilde{\bq}|+\lambda'\mu_5) \bigr] \notag\\
 &\qquad\times (1 - \lambda\hat{p}^3)(1 + \lambda'\hat{q}^3)
  \bigl[ 1 - n_{_F}(\omega_{q\lambda'})
  - \bar{n}_{_F}(\omega_{q\lambda'}) \bigr] \notag\\
 & \quad + \sum_{\lambda,\lambda'=\pm}
  \frac{\delta_{l-1,k}}
  {2\omega_{p\lambda}(\omega_{p\lambda}^2-\omega_{q\lambda'}^2)}
  \bigl[(|\tilde{\bp}|+\lambda\mu_5)^2
  +\lambda\lambda'(|\tilde{\bp}|+\lambda\mu_5)
  (|\tilde{\bq}|+\lambda'\mu_5) \bigr] \notag\\
 &\qquad\times (1 - \lambda\hat{p}^3)(1 + \lambda'\hat{q}^3)
  \bigl[ 1 - n_{_F}(\omega_{p\lambda})
  - \bar{n}_{_F}(\omega_{p\lambda}) \bigr] \,.
\label{eq:before_sum}
\end{align}
Here we note that this expression consists of two parts;  the former
part contains the distribution function with an argument
$\omega_{q\lambda'}$ (i.e.\ $n_{_F}(\omega_{q\lambda'})$ and
$\bar{n}_{_F}(\omega_{q\lambda'})$) and so the summation over
$\lambda$ can be easily taken to simplify the coefficient in front of
$[1\!-\!n_{_F}(\omega_{q\lambda'})\!-\!\bar{n}_{_F}(\omega_{q\lambda'})]$.
Remarkable simplification occurs as a result of the helicity sum,
which leads us to
\begin{equation}
 \begin{split}
 \text{Eq.~(\ref{eq:before_sum})}
  & = \sum_{\lambda'=\pm} \frac{k+1}{\omega_{q\lambda'}}
  \Bigl( 1+\frac{s'\mu_5}{|\tilde{\bq}|} \Bigr)
  \bigl[ 1-n_{_F}(\omega_{q\lambda'})
  -\bar{n}_{_F}(\omega_{q\lambda'}) \bigr] \\
 &\qquad\qquad - \sum_{\lambda=\pm} \frac{k}{\omega_{p\lambda}}
  \Bigl( 1+\frac{s\mu_5}{|\tilde{\bp}|} \Bigr)
  \bigl[ 1-n_{_F}(\omega_{p\lambda})
  -\bar{n}_{_F}(\omega_{p\lambda}) \bigr] \,.
 \end{split}
\end{equation}
From this it is apparent that there is a major cancellation in the
summation with respect to $k$ and only the edge terms remain
non-vanishing.  Hence, we carry the summation out to find the
following expression;
\begin{equation}
 \chi_{\mu_5}^\perp = VT N_c \sum_{f,s}
  \frac{q_f^2|q_f B|}{4\pi^2}\int_{-\Lambda}^\Lambda \rmd p_z
  \, \frac{\bar{k}}{\omega_{\Lambda \lambda}}
  \Bigl( 1+\frac{s\mu_5}{\Lambda} \Bigr) \bigl[
  1-n_{_F}(\omega_{\Lambda \lambda})-\bar{n}_{_F}(\omega_{\Lambda \lambda})
  \bigr] \,.
\end{equation}
Therefore, after lengthy procedures in this subsection, what we find
out at last is almost the same as Eq.~(\ref{eq:result_without_mu5})
with a minor modification by $\mu_5$, that is,
\begin{equation}
 \begin{split}
 \chi_{\mu_5}^\parallel - \chi_{\mu_5}^\perp &= VT
  N_c \sum_{f,s} \frac{q_f^2|q_f B|}{4\pi^2}
  \frac{\Lambda}{\omega_{\Lambda \lambda}} \Bigl( 1
  +\frac{s\mu_5}{\Lambda} \Bigr)
  \bigl[ 1-n_{_F}(\omega_{\Lambda \lambda})
  -\bar{n}_{_F}(\omega_{\Lambda \lambda}) \bigr] \\
 &\underset{\Lambda\to\infty}{\longrightarrow}
  VT N_c \sum_f \frac{q_f^2|q_f B|}{2\pi^2} \,.
 \end{split}
\label{eq:result_with_mu5}
\end{equation}
In the limit of $\Lambda\to\infty$ the susceptibility difference has
no dependence on $\mu_5$.


\subsection{Discussion}
\label{sec:linearresponse}

Since the final results are so simple, we will present another way to
get the same answer from heuristic arguments.  Let us consider a
situation in which a homogeneous magnetic field $\bB$ is parallel to a
homogeneous electric field $\bE$.  In such situation a current
parallel to the magnetic field will be generated by the Schwinger
process.  For infinitesimal $\bE=(0,0,E_z=E)$ the rate of change of
this current is determined by the electromagnetic anomaly relation
leading to
\begin{equation}
 \biggl\langle \frac{\rmd J_{\parallel}}{\rmd x_0} \biggr\rangle
  = V N_c \sum_f \frac{q_f^2 |q_f B| E}{2\pi^2} + O(E^2) \,.
\label{eq:djdt}
\end{equation}
We can choose a gauge so that $E=-\partial_0 A_z$, and then $A_z$ is
also infinitesimally small, which allows us to use the linear response
relation to express the current changing rate in another form,
\begin{equation}
 \biggl\langle \frac{\rmd J_{\parallel}}{\rmd x_0} \biggr\rangle
  = -\int \rmd^3 x\,\rmd^4 x'\, \Bigl\langle
  \frac{\rmd j_\parallel(x)}{\rmd x_0} j_\parallel(x')
  \Bigr\rangle_{\text{ret}} A_z(x') + O(A_z^2) \,.
\end{equation}
Assuming the translational invariance in time in
$\langle j_\parallel(x)j_\parallel(x')\rangle$, we can replace
$\rmd/\rmd x_0$ acting on it by $-\rmd/\rmd x_0'$, and then we can
perform the integration by parts to move $\rmd/\rmd x_0'$ acting onto
$A_z(x')$ which results in $E=-\partial_0 A_z$.  Eventually we arrive
at
\begin{equation}
 \biggl\langle \frac{\rmd J_{\parallel}}{\rmd x_0} \biggr\rangle
  = \int \rmd^3 x\,\rmd^4 x'\, \langle
  j_\parallel(x) j_\parallel(x') \rangle_{\text{ret}}
  E + O(E^2) \,.
\label{eq:djdt2}
\end{equation}
Identification of the left-hand side of Eq.~(\ref{eq:djdt}) with that
of Eq.~(\ref{eq:djdt2}) concludes,
\begin{equation}
 \chi^\parallel = VT N_c \sum_f \frac{q_f^2|q_f B|}{2\pi^2} \,.
\label{eq:heuristic}
\end{equation}
Now we could do the same derivation for $\langle J_\perp^2\rangle$ to
find that it is zero.  The reason is that
$\langle\rmd j_\perp/\rmd x_0\rangle$ is always vanishing for
infinitesimal $E$ in the transverse direction; small
transverse electric field cannot give rise to an electric current in
the transverse direction because of the energy barrier by the Landau
quantization.  Therefore $\chi^\perp = 0$, and so
Eq.~(\ref{eq:heuristic}) gives the difference
$\chi^\parallel - \chi^\perp$, which is in agreement with our
results~(\ref{eq:result_without_mu5}) or (\ref{eq:result_with_mu5}).

The anomaly equation is an exact relation.  Therefore, since this
derivation clearly shows that the electric-current susceptibility is
determined by the electromagnetic anomaly, we conjecture that at least
for massless quarks our results are exact and will not be modified by
including perturbative gluonic interactions that do not affect
chirality.

In Appendix~\ref{app:curchicor} we will argue that the current
susceptibility is in some sense similar to the current-chirality
correlation.  There, we will see that the anomaly relation again
constrains the current-chirality correlation.

Our results and the above-mentioned heuristic arguments are consistent
with what is known in condensed matter physics.  The hall conductivity
$\sigma_{ij}$ (the electric conductivity in the $i$-direction when
$\bE$ is imposed in the $j$-direction) in the presence of the magnetic
field is a quite familiar quantity and our $\chi^\parallel$ and
$\chi^\perp$ could be regarded as $\sigma_{zz}$ and $\sigma_{xx}$ in
the terminology of the quantum hall effect apart from the fact that
the susceptibility is a quantity in the zero-momentum limit at zero
frequency, while the conductivity can be a function of finite momentum
and frequency~\cite{Kharzeev:2009pj}.  Because our calculation has
three spatial dimensions, the counterpart in condensed matter physics
is the multilayer Dirac electron system.  Then, as discussed and
confirmed in Ref.~\cite{condensed}, only the Landau zero-mode
dominates the physics properties;  $\sigma_{xx}$ is zero because the
transverse current operator involves a shift in the Landau level as is
embodied in Eq.~(\ref{eq:delta_trans}) in our calculation, which is
actually a common knowledge in condensed matter physics.  The
longitudinal one, $\sigma_{zz}$ on the other hand, is finite and is
basically given by transport from zero-mode on one layer to zero-mode
on another layer.  In our calculation for relativistic quark matter it
is intriguing that such a finite contribution along the longitudinal
direction is uniquely constrained through the exact anomaly relation.


\section{Discussions on Lattice QCD and Experimental Data}
\label{sec:data}

Now that we have a simple expression for
$\chi_0^\parallel-\chi_0^\perp$, it would be an interesting question
whether our results are consistent with the existing results from the
lattice QCD simulation~\cite{Buividovich:2009wi} and from the previous
work~\cite{Kharzeev:2007jp} aimed at describing the CME in heavy ion
collisions.  In this section we will see that our estimate shows
reasonable agreement with lattice QCD and the formulas given in the
previous work.


\subsection{Lattice QCD Data}

In Ref.~\cite{Buividovich:2009wi} the ITEP lattice group reported on
the electric-current susceptibility in quenched QCD which they found
to be larger in the longitudinal direction than in the transverse
direction.  The simulation condition is that $N_f=1$ with $q=-e/3$,
the lattice spacing $a=0.095\fm$ (i.e.\ $\Lambda=\pi/a=6.51\GeV$), the
volume $L=16a=1.52\fm$ (and thus $V=460\GeV^{-3}$), and the simulation
is done with $N_c=2$.  Our calculations make sense above $T_c$ where
the physical degrees of freedom are quarks, so let us compare with the
lattice results at $T=1.12T_c=350\MeV$.

First of all, plugging the above numbers into Eq.~(\ref{eq:chi0}) we
find $\chi_0$ as
\begin{equation}
 \frac{\chi_0}{q^2 V^2} = \frac{2T\Lambda^2}
  {3\pi^2 V}
 = \frac{2\times 0.35\times 6.51^2}{3\pi^2\times 460}\GeV^6
 = 2.2\times 10^{-3} \GeV^6 \,,
\end{equation}
which is much larger than the lattice result.  This is not
contradictory, however.  As we have emphasized repeatedly in this
paper, a non-zero value of $\chi_0$ is an artifact of the naive
momentum cutoff, and our scheme of UV regularization is different from
the lattice one.

\begin{figure}
\begin{center}
 \includegraphics[width=0.8\textwidth]{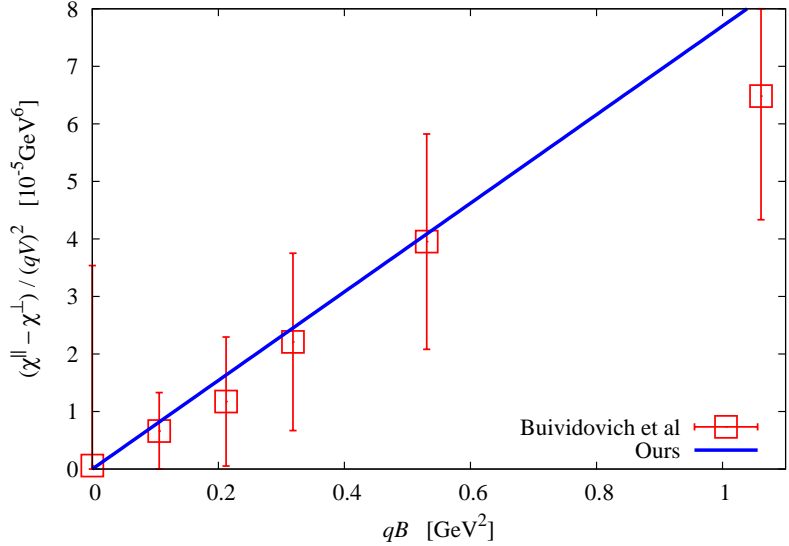}
\end{center}
 \caption{Comparison between the lattice QCD data from
   Ref.~\cite{Buividovich:2009wi} (shown by the square dots) and our
   estimate (shown by the solid line).}
 \label{fig:lattice}
\end{figure}

More interesting is the comparison of the UV-finite difference 
$\chi_0^\parallel-\chi_0^\perp$, which turns out to be
\begin{equation}
 \frac{\chi_0^\parallel - \chi_0^\perp}
  {q^2 V^2}
 = \frac{2T|qB|}{2\pi^2 V} = \frac{2\times 0.35\times |qB|}
  {2\pi^2 \times 460}
  \GeV^4 = 7.7\times 10^{-5} |qB| \GeV^4 \,.
\label{eq:ours}
\end{equation}
This is fairly close to the lattice results of 
Ref.~\cite{Buividovich:2009wi}.  To make the comparison easily visible
we shall compare our estimate with the lattice data corresponding to
Fig.~8 in Ref.~\cite{Buividovich:2009wi} for the difference between
the longitudinal and transverse susceptibilities.
Figure~\ref{fig:lattice} plots our estimate of Eq.~(\ref{eq:ours})
versus the lattice QCD data.  It is apparent that Eq.~(\ref{eq:ours})
is in good agreement within the error bars of the lattice data.


One may wonder why Eq.~(\ref{eq:ours}) works so well even though it
does not contain any information on the topological excitations and
thus no dependence on $\mu_5$.  We think that a possible explanation
is the following:  Euclidean lattice simulations do not describe the
real-time dynamics responsible for the generation of a finite $\mu_5$.
On the other hand, the Euclidean instanton transitions (that are
reproduced in lattice calculations) become suppressed in the
deconfined phase, as suggested by the rapid decrease of topological
susceptibility above $T_c$ observed on the
lattice~\cite{Vicari:2008jw}.  Because of this, the instantons may no
longer provide a major contribution to the current susceptibility in
the deconfined phase.  If so, this would provide a natural explanation
for the observation that the lattice results are almost independent of
the magnetic field at $T=1.12T_c$ \cite{Buividovich:2009wi}.  This
would also agree with the model calculation of Ref.~\cite{Nam:2009jb}
that found the CME current becoming insensitive to the magnetic field
at high temperature due to the depletion of instanton effects.  One
can test this conjecture in lattice calculations by evaluating the
susceptibility in fixed sectors of topological charge (corresponding
to a highly excited QCD vacuum configuration that arguably resembles
the matter produced in a heavy ion collision).


\subsection{Experimental Observable}

In this subsection we will check the consistency of our formula with
Ref.~\cite{Kharzeev:2007jp}.  For this purpose we shall reiterate some
phenomenological discussions in Ref.~\cite{Kharzeev:2007jp}.

Using the calculated current and susceptibility we can express the
fluctuation observables as a function of the volume $V$, the magnetic
field $B$, the chiral chemical potential $\mu_5$, etc.  That is,
\begin{align}
 & \llangle \cos(\Dphi_++\Dphi_+) \rrangle =
  -\llangle \cos(\Dphi_++\Dphi_-) \rrangle \notag\\
 &= -\frac{c}{N_\pm^2} \int\rmd\mu_5\, \W(\mu_5)\bigl(
  \langle J_\parallel \rangle_{\mu_5}^2
  + \chi_{\mu_5}^\parallel - \chi_{\mu_5}^\perp \bigr) \notag\\
 &= -\frac{c}{N_\pm^2} \int\rmd\mu_5\, \W(\mu_5)\Biggl[
  \frac{V^2 N_c^2}{4\pi^4} \Bigl(\sum_f q_f^2 \Bigr)^2 B^2\mu_5^2
  + VT N_c \sum_f \frac{q_f^2|q_f B|}{2\pi^2} \Biggr] \,.
\label{eq:av_cos}
\end{align}
The problem is that this expression at first sight may appear
different from that discussed previously in
Ref.~\cite{Kharzeev:2007jp}.  However below we explain that the first
term stemming from the CME in fact has essentially the same form as
discussed in Ref.~\cite{Kharzeev:2007jp}.

In describing the real data, it is necessary to take account of the
screening effect in the $z$ direction (denoted previously as $y$ in
Ref.~\cite{Kharzeev:2007jp}).  Because $\mu_5$ emulates the effect of
topologically non-trivial domains in a hot medium, it would be a
reasonable interpretation that our estimate is a valid answer in each
domain with a finite volume $V\simeq\rho^3$.  Here, if the topological
domain is characterized by a sphaleron excitation,
$\rho\sim 1/(\alpha_s T)$ is the typical sphaleron size.

Now the sphaleron can get excited anywhere in the medium and we should
sum over all excitations.  If an excitation occurs deep in a medium,
it is unlikely that current fluctuations from the corresponding domain
can propagate to survive outside because of medium screening.  On the
other hand a sphaleron excitation near the surface can easily escape
from the medium, and then it should contribute to the total current
fluctuations if the correlator is for same charges.  In the case of
opposite charges, however, as is clear from our discussions in
Sec.~\ref{sec:observable}, fluctuations arise from, for example, an
upgoing positive charge and a downgoing negative charge.  Thus, one of
two charges must be significantly quenched by the medium.  In
Ref.~\cite{Kharzeev:2007jp} a phenomenological ansatz for such effects
was introduced by the following functions;
\begin{align}
 g(b/R,\lambda/R) &\equiv \frac{1}{2R^2}\int_{-R+b/2}^{R-b/2} \rmd x
  \int_{z_-(x)}^{z_+(x)} \rmd z\,
  \bigl[\, \xi_+^2(x,z)+\xi_-^2(x,z) \bigr] \,, \\
 h(b/R,\lambda/R) &\equiv \frac{1}{R^2}\int_{-R+b/2}^{R-b/2} \rmd x
  \int_{z_-(x)}^{z_+(x)} \rmd z\;
  \xi_+(x,z)\,\xi_-(x,z) \,,
\end{align}
for the correlations with same and opposite charges, respectively,
where
\begin{equation}
 \xi_\pm(x,z) \equiv \exp\bigl[ -|z_\pm(x)-z|/\lambda \bigr]
\end{equation}
with a phenomenological screening length $\lambda$ and
\begin{equation}
 z_+(x) = -z_-(x) = \left\{
  \begin{array}{lp{1em}l}
  \sqrt{R^2-(x-b/2)^2\,} && -R+b/2 \le x \le 0 \\
  \sqrt{R^2-(x+b/2)^2\,} && 0 \le x \le R-b/2
  \end{array} \right.
\end{equation}
is the surface of matter.  Here $b$ is the impact parameter, $R$ is
the radius of the nucleus.

We postulate that the superposition of the sphaleron domains with $V$
over the whole system geometry amounts to the system volume factor
with the quenching effect taken into account, that is, $V$ is replaced
as
\begin{equation}
 V\sum_{\text{domains}} \longrightarrow \left\{ \begin{array}{lp{1em}l}
 V_{++} \equiv g(b/R,\lambda/R) R^2 \Delta\eta\,\tau &&
  \text{ (for $\pm\pm$ correlations) } \\
 V_{+-} \equiv h(b/R,\lambda/R) R^2 \Delta\eta\,\tau &&
  \text{ (for $\pm\mp$ correlations) }
 \end{array} \right. \,.
\end{equation}
At this point let us think of the $\mu_5$-integration.  The latter
term in Eq.~(\ref{eq:av_cos}) does not have any $\mu_5$ dependence, so
the $\mu_5$-integration is trivial leading to
$\int\!\rmd\mu_5\W(\mu_5)=1$.  The former term, in contrast, is
proportional to $\mu_5^2$.  This means that $\mu_5^2$ after the
integration turns to be a parameter (denoted as $\mu_0$ here)
characterizing the dispersion of the $\mu_5$ distribution.
Consequently we have the final expressions for the respective cases
with the same and opposite charges;
\begin{equation}
 \llangle \cos(\Dphi_\pm+\Dphi_\pm) \rrangle
 = -\frac{c\,V_{++}}{N_\pm^2} \Biggl[ \,
  \frac{N_c^2\rho^3}{4\pi^4} \Bigl(\sum_f q_f^2\Bigr)^2 B^2\mu_0^2
  + T N_c\sum_f \frac{q_f^2|q_f B|}{2\pi^2} \Biggr] \,,
\label{eq:final_result++}
\end{equation}
and $\llangle \cos(\Dphi_\pm+\Dphi_\mp) \rrangle$ given by almost the
same with $V_{++}$ replaced by $V_{+-}$ in the above.  Here we note
that we should identify $\mu_0^2$ as
\begin{equation}
 \mu_0^2 \propto \Gamma \Delta\tau \rho \,,
\end{equation}
where $\Gamma$ is the sphaleron rate which is proportional to
$\alpha_s^5T^4$ at high $T$ and thus $\Gamma\Delta\tau$ represents how
much topological excitations occur within a time slice $\Delta\tau$.
By the dimensional reason $\rho$ appears.  Then we clearly see that
the first term in Eq.~(\ref{eq:final_result++}) has exactly the same
structure as the expression discussed in Ref.~\cite{Kharzeev:2007jp}
apart from its overall coefficient once it is integrated over the
time; $\Delta\tau\to\int\!\rmd\tau$.  In this way we have established
a relation between the formalism addressed here and the previous work
in Ref.~\cite{Kharzeev:2007jp}.

To proceed to more concrete analysis we need to specify $\lambda$,
$B(\tau)$ and $N_\pm$ as a function of $b$ (or the centrality).
Although those phenomenological analyses are important, we will
postpone such investigations and discuss them in a separate
publication which focuses more on the phenomenology of heavy-ion
collisions.


\section{Summary}
\label{sec:summary}

In this paper we formulated the Chiral Magnetic Effect in terms of the
electric-current correlation function in a hot QCD medium.  We
computed the electric-current susceptibility in the presence of both
the magnetic field $B$ and the chiral chemical potential $\mu_5$.
Because of the presence of a preferred direction fixed by $\bB$, we
found that the longitudinal (parallel to $B$) susceptibility
$\chi^\parallel$ is greater than the transverse (perpendicular to $B$)
one $\chi^\perp$.  The difference arises from only the Landau
zero-mode and is given by a UV-finite expression;
$VTN_c\sum_f q_f^2|q_f B|/(2\pi^2)$.  We also gave an intuitive
derivation of our result based on the anomaly relation.  We checked
that our result leads to a satisfactory agreement with the
electric-current susceptibility measured in the lattice QCD simulation
\cite{Buividovich:2009wi}.

Although $\chi^\parallel-\chi^\perp$ has an origin in the anomaly, the
expression for $\chi^\parallel-\chi^\perp$ shows that this difference is
independent of $\mu_5$ and thus is not sensitive to the real-time
topological contents of the QCD matter.  Therefore we should identify
it as a background on top of the CME contribution stemming from the
square of the induced CME current.  Since the CME-induced current is
proportional to $B$ and $\mu_5$, the charge-asymmetry fluctuation
relevant to the CME has a dependence of $B^2$ and $\mu_5^2$ which
translates into a dispersion parameter of the $\mu_5$-distribution; on
the other hand, the non-CME term is proportional to $B$.  It would be
an interesting question to check if this expected quadratic dependence
on $B$ can be seen in a lattice simulation restricted to a particular
topological sector of QCD, that would correspond to a highly excited
vacuum configuration resembling the matter produced in a heavy ion
collision.

There is an urgent need at the moment in the quantitative
phenomenological approaches to CME that on one hand have a firm
theoretical ground and on the other hand allow a direct comparison to
the experimental data from RHIC~\cite{:2009uh}.  We view the
computation presented here as a necessary step towards a construction
of such an approach.  We will report on our approach and on the direct
quantitative description of RHIC data in the forthcoming separate
publication.


\section*{Acknowledgments}
We thank Pavel Buividovich, Maxim Chernodub, and Mikhail Polikarpov
for useful discussions and for kindly providing us with their lattice
QCD data.  We are grateful to Tom Blum for valuable discussions on CME
with light quarks.  K.~F.\ thanks Takao Morinari for discussions in
relation to condensed matter physics.  The work of K.~F.\ was
supported by Japanese MEXT grant No.\ 20740134 and also supported in
part by Yukawa International Program for Quark Hadron Sciences.  The
work of D.~K.\ was supported by the Contract No.\ \#DE-AC02-98CH10886
with the U.S.\ Department of Energy.  The work of H.J.~W.\ was
supported by the Alexander von Humboldt Foundation.

\appendix

\section{Appendix: Induced current from the diagrammatic method}
\label{app:curprop}

In Refs.~{\cite{Fukushima:2008xe}} and {\cite{Kharzeev:2009pj}} we
have discussed several different derivations of the induced current in
a magnetic field in the presence of non-zero chirality.  Here we add
an alternative derivation using the propagator
Eq.~(\ref{eq:propagator}).

From Eq.~(\ref{eq:propagator}) we readily obtain for the induced
current along the magnetic field,
\begin{equation}
 \begin{split}
\langle J_\parallel \rangle_{\mu_5} &= 
 \rmi\, V N_c
\sum_f \frac{q_f \vert q_f B\vert }{2\pi} \sum_k 
\int^T \frac{\rmd p_0}{2\pi} \int 
  \frac{\rmd p_z}{2\pi} 
\int \rmd x 
\\ &\qquad\qquad \times
 \tr \Bigl[ \gamma^3 
 P_k(x)\, \bigl(\feyn{\tilde{p}} + \mu_5 \gamma^0 \gamma^5
 - M_f \bigr)^{-1} P_k (x) \Bigr] 
\,.
 \end{split}
\label{eq:curprop1}
\end{equation}
Since $\gamma^3$ commutes with $P_k(x)$ we need to evaluate the
integral over $x$ of $P_k(x)^2$ which equals $1$ for $k>0$ and
$(1+ \rmi\, \sgn(q_f B) \gamma^1 \gamma^2)/2$ for $k=0$. Inserting the
projection operators
$\Gamma_\pm(\tilde p)$ like in Eq.~(\ref{eq:suscepmu5}) and using
Eq.~(\ref{eq:diracoperatorinversion}) it can be seen that we need to
evaluate the following two traces;
\begin{align}
 & \tr \Bigl[\gamma^3 
\bigl(
\feyn{\tilde{p}} + \mu_5 \gamma^0 \gamma^5 + M_f 
\bigr)
\Gamma_\pm(\tilde p)
 \Bigr] =
2p_z \Bigl( 1 \pm \frac{\mu_5}
{\vert \tilde{\bp}\vert} \Bigr) \,,
\\
 & \rmi\, \tr \Bigl[\gamma^1 \gamma^2 \gamma^3 
\bigl(
\feyn{\tilde{p}} + \mu_5 \gamma^0 \gamma^5 + M_f 
\bigr)
\Gamma_\pm(\tilde p)
 \Bigr]
=
-2 \mu_5 \mp 2 \vert \tilde{\bp} \vert \,.
\end{align}
For $k > 0$ we only need the former trace, which vanishes after
integration over $p_z$.  The only contribution to the current comes
from the latter trace when $k=0$, which is the lowest Landau level.
Hence we obtain after summing over Matsubara frequencies,
\begin{equation}
\langle J_\parallel \rangle_{\mu_5} = 
 V N_c
\sum_f \frac{q_f^2 B}{2\pi} 
\int 
  \frac{\rmd p_z}{2\pi} \sum_{\lambda=\pm}
\frac{\mu_5 + \lambda \vert p_z \vert}{2\omega_{p\lambda}}
\bigl[
1 - n_{_F}(\omega_{p\lambda}) - \bar{n}_{_F}(\omega_{p\lambda})
\bigr]
\,,
\label{eq:curprop2}
\end{equation}
where $\omega_{p\lambda} = \sqrt{( |p_z| + \lambda \mu_5)^2 + M_f^2}$
for the Landau zero-mode.  Noticing that
$(\mu_5+\lambda|p_z|)/\omega_{p\lambda}
=\lambda\,\rmd\omega_{p\lambda}/\rmd p_z$ and integrating over $p_z$
we obtain like in Eq.~(\ref{eq:curmu5}) the known result;
$\langle J_\parallel \rangle_{\mu_5}
= V N_c \sum_f q_f^2 B \mu_5 / (2\pi^2)$, which is independent of
$M_f$, $\mu_q$ and $T$.

\section{Appendix: Current-chirality correlation}
\label{app:curchicor}

Here we will discuss the correlation of the chirality with the current
in the direction parallel to the magnetic field.  We will show that
the magnitude of this correlation is very similar to the difference
between the longitudinal and transverse susceptibility.

By taking the derivative of Eq.~(\ref{eq:curmu5}) with respect to
$\mu_5$ one can easily find the correlation between the chirality and
the  longitudinal current.  Such a calculation immediately yields,
\begin{equation}
 \langle J_{\parallel}\, N_5 \rangle_{\text{connected}} =
 V T N_c \sum_f \frac{q_f^2 B}{2 \pi^2} \,,
\label{eq:curn5cor}
\end{equation}
where the total chirality $N_5$ is the volume integral over the
zero-component of the axial charge, i.e.\
$N_5 = \sum_f \int\! \rmd^3 x
\langle \bar\psi_f \gamma^0 \gamma^5 \psi_f \rangle$.  This
correlation function shows that in a magnetic field the longitudinal
current is correlated with the chirality, which is the Chiral Magnetic
Effect.  The magnitude of  this correlation is very similar to the
difference  between the longitudinal and transverse susceptibility as
can be inferred from Eq.~(\ref{eq:result_with_mu5}). The only
alteration comes from the charges $q_f$ because the chirality is not
accompanied by the charge unlike the current.  However,
$\langle J_\parallel\, N_5\rangle_{\text{connected}}$ depends on the
sign of $B$, while $\chi^\parallel_{\mu_5} - \chi^\perp_{\mu_5}$ does
not.

In the same way as in Sec.~\ref{sec:linearresponse} we could have also
arrived at Eq.~(\ref{eq:curn5cor}) using the linear response theory.
In that case we start from the anomaly equation for massless particles
which reads
\begin{equation}
 \biggl\langle \frac{ \rmd N_5}{\rmd x_0} \biggr\rangle = 
 V N_c \sum_f \frac{q_f^2 E B}{2\pi^2} \,.
\label{eq:dn5dt}
\end{equation}
Applying the same arguments as in Sec.~\ref{sec:linearresponse} and
replacing one $J_\parallel$ with $N_5$ in the arguments we can recover
Eq.~(\ref{eq:curn5cor}) correctly.

The reason why the current-chirality correlator is so similar to the
difference between the longitudinal and transverse susceptibility is
that both quantities follow from the axial anomaly.  This can be
understood explicitly in view of the respective starting points of the
linear response derivation, namely, Eq.~(\ref{eq:djdt}) and
Eq.~(\ref{eq:dn5dt}).


\end{document}